\documentclass[prc,letterpaper,showpacs,12pt,floatfix,preprint]{revtex4}

\usepackage{amsmath}
\usepackage{bm}
\usepackage{graphicx}

\begin{document}

\preprint{\vbox{\hbox{JLAB-THY-08-822}}}

\title{A new calculation for $D(e,e'p)n$ at GeV energies}

\author{Sabine Jeschonnek$^{(1)}$ and J. W. Van Orden$^{(2,3)}$}

\affiliation{\small \sl (1) The Ohio State University, Physics
Department, Lima, OH 45804\\
(2) Department of Physics, Old Dominion University, Norfolk, VA
23529\\and\\ (3) Jefferson Lab\footnote{Notice: Authored by Jefferson Science Associates, LLC under U.S. DOE Contract No. DE-AC05-06OR23177. The U.S. Government retains a non-exclusive, paid-up, irrevocable, world-wide license to publish or reproduce this manuscript for U.S. Government purposes}, 12000 Jefferson Avenue, Newport
News, VA 23606
 }

\date{\today}

\begin{abstract}
We perform a fully relativistic calculation of the $D(e,e'p)n$ reaction in the impulse approximation. We employ the Gross equation
to describe the deuteron ground state, and we use the SAID parametrization of the full NN scattering amplitude
to describe the final state interactions (FSIs). We include both on-shell and positive-energy off-shell contributions
in our FSI calculation. We show results for momentum distributions and angular distributions of the
differential cross section, as well as for various asymmetries. We identify kinematic regions where various
parts of the final state interactions are relevant, and discuss the theoretical uncertainties connected
with calculations at high missing momenta.

\end{abstract}
\pacs{25.30.Fj, 21.45.Bc, 24.10.Jv}

\maketitle

\section{Introduction}

Exclusive electron scattering from the deuteron target is very
interesting by itself, and also as a very relevant stepping
stone towards understanding exclusive electron scattering from
heavier nuclei. The $D(e,e'p)n$ reaction at GeV energies allows us -
and requires us - to carefully study the reaction mechanism. It is
necessary to consider final state interactions (FSIs) between the
two nucleons in the final state, two-body currents, and isobar
contributions. Of these, the FSIs can be expected to be the most
relevant part of the reaction mechanisms at the GeV energy and
momentum transfers relevant to the study of the transition from
hadronic to quark-gluon degrees of freedom. For some recent reviews
on this exciting topic, see e.g. \cite{wallyreview,ronfranz,sickreview}.

Even though the deuteron is the simplest nucleus, and it has been
the subject of considerable attention for a long time, there are
several open questions: Is it possible to experimentally determine the high-momentum components of the deuteron wave function even though the nuclear wave function or the momentum distribution
are not observables? Will conventional nuclear physics break
down or become too cumbersome at some point, and will a description
involving quarks become necessary? Are there any six-quark
admixtures in the deuteron wave function, and do they have an
unambiguous experimental signal? What influence does short range
physics have on conventional wave functions, and how can this
influence be removed to find an effective potential $V_{low \, k}$,
provided one is interested in low energy scenarios only
\cite{srgosu,roth}? Relativistic wave functions are available
\cite{wallyfranzwf} for the deuteron. And while the calculational
effort is still considerable, it can be managed without resorting to
super computers.
The interest and importance of the
$D(e,e'p)n$ process is reflected in the fact that recently, a
deuteron benchmarking project has been started to investigate the
differences between various calculations that are based on
non-relativistic wave functions \cite{dbm}.

Anything that we learn about the reaction mechanism of the
$D(e,e'p)n$ reaction has implications for heavier targets or
experiments where the deuteron is used as a lab. Some examples for
the latter are the measurement of the neutron magnetic form factor
by measuring a ratio of $D(e,e'p)$ and $D(e,e'n)$ cross sections.
This allows for a significant reduction in the model dependence of
the extracted form factor, but some theoretical input is still
required \cite{hallbgmn}. Another example for using the nucleus
as a lab is color transparency. While meson production from nuclei
\cite{haiyanct,ctjan,ctmiller,ctlaget} is the main thrust of color
transparency investigations, color transparency in $(e,e'p)$
reactions is very interesting and topical, too \cite{cteep}. In
order to study color transparency, one first needs to establish a
firm understanding of all the conventional nuclear effects.

Several experiments on deuteron targets have been performed
in the last few years, both at Jefferson Lab and at MIT Bates,
and these data have either been published recently or are currently under analysis
\cite{egiyansrc,halladata,hallbgmn,jerrygexp,blast}.
There are also new proposals for D(e,e'p) experiments at Jefferson Lab \cite{wernernewprop}.
Apart from these exciting new data, there are very interesting open questions posed by the
data that have been available for some years now \cite{paulprl,bernheim,boeglintrento2005,voutier}. Regardless of the momentum
transfers involved, there has been a discrepancy between data and calculations at low missing momentum.
We discuss calculations in kinematics relevant to the new experiments as well as the low missing momentum
puzzle in our results section.

The experimental activity in $(e,e'p)$ reactions on the deuteron and other light nuclei has been matched by theoretical
efforts. These calculations typically are performed using Glauber theory, the generalized eikonal approximation
\cite{misak,ciofi,genteikonal}, or a diagrammatic approach \cite{laget}, although there are rare exceptions \cite{schiavilla}.
Even a second-order correction to the eikonal approximation has been suggested recently \cite{secondordereikonal}.
Many of these calculations focus on the differential
cross section only, and use just the central part of the NN scattering amplitude. Currently, almost all calculations for $D(e,e'p)n$ reactions
are unfactorized  \cite{sabinefac}, but factorized approaches are used for heavier targets \cite{ciofifactor}.
A common feature first introduced in Glauber calculations is the assumption that the momentum transfer in the
rescattering of the two nucleons is purely transverse. This has consequences both for the profile function, and the
argument for which the NN scattering amplitude is evaluated.

In this paper, we present a new calculation with several important features: we use a fully relativistic formalism,
and describe the ground state with a solution of the Gross equation\cite{grosseqn}; we include all parts of the nucleon-nucleon
scattering amplitude, including all the spin dependent parts, and use a realistic, modern parametrization;
the only approximation that we make is to neglect the negative energy states, as discussed at the
end of the first section. This new calculation
can be used at all energy and momentum transfers, provided that an appropriate full set of $pn$ scattering data
is available. This paper is organized as follows: First, we review the theoretical framework for our calculations, then, we show numerical results for the cross section and  asymmetries. We conclude with a summary and outlook.

\section{Theoretical Framework}
\label{sectheo}

\begin{figure}
\centerline{\includegraphics[height=2.5in]{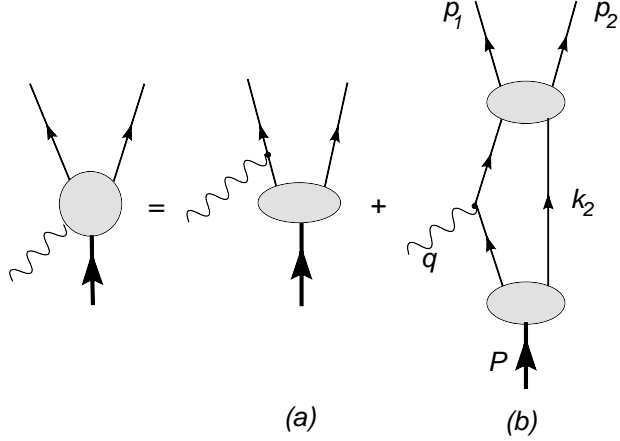}}
\caption{Feynman diagrams representing the impulse
approximation.  }\label{impulse}
\end{figure}

The Feynman diagrams representing the impulse approximation are
shown in Fig. \ref{impulse}.
Figure \ref{impulse}a represents the plane wave contribution and
Fig. \ref{impulse}b represents the contribution from final state
interactions. The plane wave contribution to the current matrix
element is given by
\begin{equation}
\left<\bm{p}_1s_1;\bm{p}_2s_2\right|J^\mu_{PW}\left|\bm{P}\lambda_d\right>
=-\bar{u}(\bm{p}_1,s_1)\Gamma^\mu(q)G_0(P-p_2)
\Gamma^T_{\lambda_d}(p_2,P)\bar{u}^T(\bm{p}_2,s_2)\,,
\end{equation}
where the target deuteron has four-momentum $P$ and spin
$\lambda_d$, the final proton has four-momentum $p_1$ and spin $s_1$
and the final neutron has four-momentum $p_2$ and spin $s_2$. The
single-nucleon propagator is
\begin{equation}
G_0(p)=\frac{\gamma\cdot p+m}{m^2-p^2-i\eta}
\end{equation}
and the current operator is chosen to be of the free
Dirac-plus-Pauli form
\begin{equation}
\Gamma^\mu(q)=F_1(Q^2)\gamma^\mu+\frac{F_2(Q^2)}{2m}i\sigma^{\mu\nu}q_\nu\,.
\end{equation}
The deuteron vertex function with nucleon 2 on shell can be written
as
\begin{eqnarray}
\Gamma_{\lambda_d}(p_2,P)&=&g_1(p_2^2,p_2\cdot
P)\gamma\cdot\xi_{\lambda_d}(P) +g_2(p_2^2,p_2\cdot P)
\frac{p\cdot\xi_{\lambda_d}(P)}{m}\nonumber\\
&&-\left(g_3(p_2^2,p_2\cdot P)\gamma\cdot\xi_{\lambda_d}(P)
+g_4(p_2^2,p_2\cdot
P)\frac{p\cdot\xi_{\lambda_d}(P)}{m}\right)\frac{\gamma\cdot
p_1+m}{m}C\,,
\end{eqnarray}
where $p_1=P-p_2$, $p=\frac{1}{2}(p_1-p_2)=\frac{P}{2}-p_2$, $C$ is
the charge-conjugation matrix and $\xi_{\lambda_d}$ is the deuteron
polarization four-vector. The invariant functions $g_i$ are given by
\begin{eqnarray}
g_1(p_2^2,p_2\cdot P)&=&\frac{2E_k-M_d}{\sqrt{8\pi}}\left[ u(k)-\frac{1}{\sqrt{2}}w(k)+\sqrt{\frac{3}{2}}\frac{m}{k}v_t(k)\right]\\
g_2(p_2^2,p_2\cdot P)&=&\frac{2E_k-M_d}{\sqrt{8\pi}}\left[ \frac{m}{E_k+m}u(k)+\frac{m(2E_k+m)}{\sqrt{2}k^2}w(k)+\sqrt{\frac{3}{2}}\frac{m}{k}v_t(k)\right]\\
g_3(p_2^2,p_2\cdot P)&=&\sqrt{\frac{3}{16\pi}}\frac{m E_k}{k}v_t(k)\\
g_4(p_2^2,p_2\cdot P)&=&-\frac{m^2}{\sqrt{8\pi}M_d}\left[(2 E_k-M_d)\left(\frac{1}{E_k+m}u(k) -\frac{E_k+2m}{\sqrt{2}k^2}w(k)\right)+\frac{\sqrt{3}M_d}{k}v_s(k)\right]\,,\nonumber\\
\end{eqnarray}
where
\begin{equation}
k=\sqrt{\frac{(P\cdot p_2)^2}{P^2}-p_2^2}
\end{equation}
is the magnitude of the neutron three-momentum in the deuteron rest
frame and
\begin{equation}
E_k=\sqrt{k^2+m^2}\,.
\end{equation}
The functions $u(k)$, $w(k)$, $v_s(k)$ and $v_t(k)$ are the s-wave,
d-wave, singlet p-wave and triple p-wave radial wave functions of
the deuteron in momentum space.  The radial wave functions are
normalized in the absence of energy-dependent kernels such that
\begin{equation}
\int_0^\infty
\frac{dpp^2}{(2\pi)^3}\left[u^2(p)+w^2(p)+v_t^2(p)+v_s^2(p)\right]=1\,.
\end{equation}
For convenience, the spectator deuteron wave function can be defined
as
\begin{equation}
\psi_{\lambda_d,s_2}(p_2,P)=G_0(P-p_2)
\Gamma^T_{\lambda_d}(p_2,P)\bar{u}^T(\bm{p}_2,s_2)\,.
\end{equation}
We choose to normalize this wave function such that in the deuteron
rest frame
\begin{equation}
\sum_{s_2}\int\frac{d^3p_2}{(2\pi)^3}\frac{m}{E_{p_2}}\bar{\psi}_{\lambda_d,s_2}(p_2,P)
\gamma^0 \psi_{\lambda_d,s_2}(p_2,P)=1\,,
\end{equation}
which is correct only in the absence of energy-dependent kernels.
The plane wave contribution to the current matrix element can then
be written as
\begin{equation}
\left<\bm{p}_1s_1;\bm{p}_2s_2\right|J^\mu_{PW}\left|\bm{P}\lambda_d\right>=
-\bar{u}(\bm{p}_1,s_1)\Gamma^\mu(q)\psi_{\lambda_d,s_2}(p_2,P)
\end{equation}

\begin{figure}
\centerline{\includegraphics[height=2.5in]{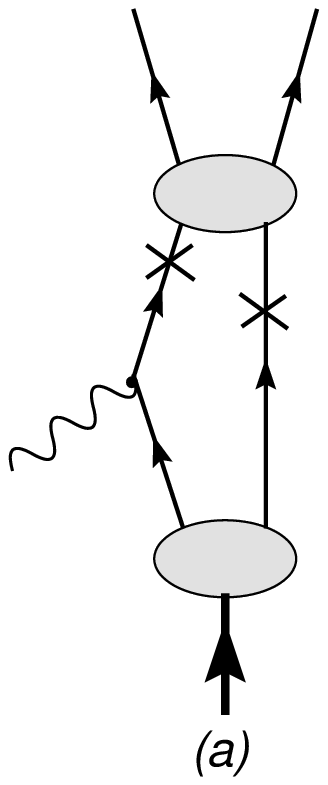}\includegraphics[height=2.5in]{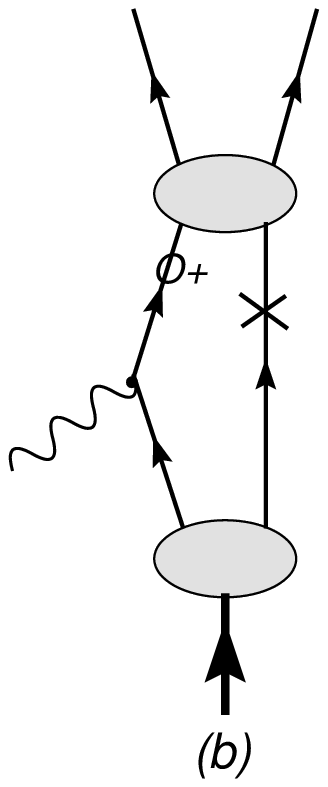}\includegraphics[height=2.5in]{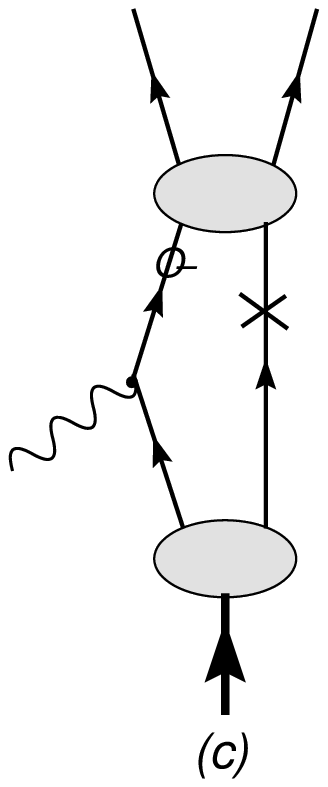}}
\caption{Diagrams representing
$\left<\bm{p}_1s_1;\bm{p}_2s_2\right|J^\mu_{a}\left|\bm{P}\lambda_d\right>$,
$\left<\bm{p}_1s_1;\bm{p}_2s_2\right|J^\mu_{b}\left|\bm{P}\lambda_d\right>$
and
$\left<\bm{p}_1s_1;\bm{p}_2s_2\right|J^\mu_{c}\left|\bm{P}\lambda_d\right>$.
The insertion of a cross on an internal propagator indicates that
the particle has been placed on the positive-energy mass shell. The
insertion $O+$ indicates that the positive-energy principal value
part of the propagator is used and the insertion of $O-$ indicates
the negative-energy part of the propagator is used.  }\label{Jabc}
\end{figure}

The contribution from final state interactions represented by Fig.
\ref{impulse}b requires an integration for the loop four-momentum
$k_2$ which involves both the deuteron vertex function and the $pn$
scattering amplitude. An equivalent approach is to formulate the
problem in terms of the Spectator or Gross equations \cite{grosseqn} where the
equations for the scattering amplitude and vertex function are
rewritten such that one particle is always taken to be on mass
shell. This approach is manifestly covariant and has been
successfully applied to elastic electron scattering from the
deuteron \cite{wallyfranz}. Using this approach, the contribution of the final state
interaction to the current matrix element is given by
\begin{eqnarray}
\left<\bm{p}_1s_1;\bm{p}_2s_2\right|J^\mu_{FSI}\left|\bm{P}\lambda_d\right>&=&\int
\frac{d^3k_2}{(2\pi)^3}\frac{m}{E_{k_2}}
\bar{u}_a(\bm{p}_1,s_1)\bar{u}_b(\bm{p}_2,s_2)M_{ab;cd}(p_1,p_2;k_2)
\nonumber\\
&&\times{G_0}_{ce}(P+q-k_2)\Gamma^\mu_{ef}(q){G_0}_{fg}(P-k_2)\nonumber\\
&&\times\Lambda^+_{dh}(\bm{k}_2){\Gamma^T_{\lambda_d}}_{gh}(k_2,P)\,,\label{J_FSI}
\end{eqnarray}
where $M$ is the $pn$ scattering amplitude,
\begin{equation}
\Lambda^+(\bm{p})=\sum_s
u(\bm{p},s)\bar{u}(\bm{p},s)=\frac{\gamma\cdot p+m}{2m}
\end{equation}
is the positive energy projection operator and the Dirac indices for
the various components are shown explicitly.  Since the
single-nucleon propagator can be decomposed as
\begin{eqnarray}
G_0(p)&=&-\frac{m}{E_p}\sum_s\left[\frac{u(\bm{p},s)\bar{u}(\bm{p},s)}{p^0-E_p+i\epsilon}
+\frac{v(-\bm{p},s)\bar{v}(-\bm{p},s)}{p^0+E_p-i\epsilon}\right]\nonumber\\
&=&-\frac{m}{E_p}\left[\frac{\Lambda^+(\bm{p})}{p^0-E_p+i\epsilon}
-\frac{\Lambda^-(-\bm{p})}{p^0+E_p-i\epsilon}\right]
\label{g0ofp}
\end{eqnarray}
and
\begin{equation}
\frac{1}{p^0-E_p+i\epsilon}=-i\pi\delta(p^0-E_p)+\frac{\cal{P}}{p^0-E_p}\,,
\end{equation}
(\ref{J_FSI}) can be written as
\begin{equation}
\left<\bm{p}_1s_1;\bm{p}_2s_2\right|J^\mu_{FSI}\left|\bm{P}\lambda_d\right>=
\left<\bm{p}_1s_1;\bm{p}_2s_2\right|J^\mu_a
\left|\bm{P}\lambda_d\right>
+\left<\bm{p}_1s_1;\bm{p}_2s_2\right|J^\mu_b\left|\bm{P}\lambda_d\right>
+\left<\bm{p}_1s_1;\bm{p}_2s_2\right|J^\mu_c\left|\bm{P}\lambda_d\right>
\end{equation}
where
\begin{eqnarray}
\left< \bm{p}_1s_1;\bm{p}_2s_2\right|J^\mu_{a}\left|\bm{P}\lambda_d\right>&=&i\pi\sum_{\sigma_2}\int \frac{d^3k_2}{(2\pi)^3}\frac{m}{E_{k_2}}\frac{m}{E_{P+q-k_2}}\delta(P^0+\nu-E_{k_2}-E_{P+q-k_2})\nonumber\\
&&\qquad\times\bar{u}_a(\bm{p}_1,s_1)\bar{u}_b(\bm{p}_2,s_2)M_{ab;cd}(p_1,p_2;k_2)u_d(\bm{k}_2,\sigma_2)
\nonumber\\
&&\qquad\times\left(\Lambda^+(\bm{P}+\bm{q}-\bm{k}_2)\Gamma^\mu(q)\psi_{\lambda_d,\sigma_2}(k_2,P)\right)_c\,,
\label{J_FSIa}
\end{eqnarray}

\begin{eqnarray}
\left<\bm{p}_1s_1;\bm{p}_2s_2\right|J^\mu_{b}\left|\bm{P}\lambda_d\right>&=&-\sum_{\sigma_2}{\cal{P}}\int \frac{d^3k_2}{(2\pi)^3}\frac{m}{E_{k_2}}\frac{m}{E_{P+q-k_2}}\frac{1}{P^0+\nu-E_{k_2}-E_{P+q-k_2}}\nonumber\\
&&\qquad\times\bar{u}_a(\bm{p}_1,s_1)\bar{u}_b(\bm{p}_2,s_2)M_{ab;cd}(p_1,p_2;k_2)u_d(\bm{k}_2,\sigma_2)
\nonumber\\
&&\qquad\times\left(\Lambda^+(\bm{P}+\bm{q}-\bm{k}_2)\Gamma^\mu(q)\psi_{\lambda_d,\sigma_2}(k_2,P)\right)_c
\label{J_FSIb}
\end{eqnarray}
and
\begin{eqnarray}
\left<\bm{p}_1s_1;\bm{p}_2s_2\right|J^\mu_{c}\left|\bm{P}\lambda_d\right>&=&\sum_{\sigma_2}\int
\frac{d^3k_2}{(2\pi)^3}\frac{m}{E_{k_2}}\frac{m}{E_{P+q-k_2}}\frac{1}{P^0+\nu-E_{k_2}+E_{P+q-k_2}
-i\epsilon}\nonumber\\
&&\qquad\times\bar{u}_a(\bm{p}_1,s_1)\bar{u}_b(\bm{p}_2,s_2)M_{ab;cd}(p_1,p_2;k_2)u_d(\bm{k}_2,\sigma_2)
\nonumber\\
&&\qquad\times\left(\Lambda^-(
\bm{k}_2-\bm{q}-\bm{P})\Gamma^\mu(q)\psi_{\lambda_d,\sigma_2}(k_2,P)\right)_c\,.\label{J_FSIc}
\end{eqnarray}
These contributions are represented by the diagrams in Fig.
\ref{Jabc}.

Equation (\ref{J_FSIa}), represented by  Fig. \ref{Jabc}a, has all four
legs of the $pn$ scattering amplitude on mass shell. For this case,
the scattering amplitude can be parameterized in terms of five Fermi
invariants as
\begin{eqnarray}
M_{ab;cd}&=&\mathcal{F}_S(s,t)\delta_{ac}\delta_{bd}+\mathcal{F}_V(s,t)\gamma_{ac}
\cdot\gamma_{bd}+\mathcal{F}_T(s,t)\sigma^{\mu\nu}_{ac}(\sigma_{\mu\nu}) _{bd}^{} \nonumber\\
&&+\mathcal{F}_{P}(s,t)\gamma^5_{ac}\gamma^5_{bd}+
\mathcal{F}_A(s,t)(\gamma^5\gamma)_{ac}\cdot(\gamma^5\gamma)_{bd}\label{Fermi}
\label{eqdefnn}
\end{eqnarray}
where $s$ and $t$ are the usual Mandelstam variables. The
calculation of the invariant functions from helicity amplitudes is
described in Appendix \ref{appnn}. We use the helicity amplitudes
available from SAID as input for our calculations \cite{said}.
For this calculation we have constructed
a table of the invariant functions in terms of $s$ and the center of
momentum angle $\theta$.  The table is then interpolated to obtain the
invariant functions at the values required by the integration.

An alternative two-dimensional representation of the scattering
amplitudes is in terms of the Saclay amplitudes.  In this case, the
scattering amplitude as an operator in two-dimensional spinor space
is given by
\begin{eqnarray}
\widetilde{M}&=&\frac{1}{2}\left[(a_s+b_s)
+(a_s-b_s)\bm{\sigma}_1\cdot\hat{\bm{n}}\bm{\sigma}_2\cdot\hat{\bm{n}}
+(a_s-b_s)\bm{\sigma}_1\cdot\hat{\bm{m}}\bm{\sigma}_2\cdot\hat{\bm{m}}\right.\nonumber\\
&&\left.+(a_s-b_s)\bm{\sigma}_1\cdot\hat{\bm{l}}\bm{\sigma}_2\cdot\hat{\bm{l}}
+e_s(\bm{\sigma}_1+\bm{\sigma}_2)\cdot\hat{\bm{n}}\right]
\label{saclaynndef}
\end{eqnarray}
where
\begin{equation}
\hat{\bm{l}}=\frac{\bm{p}'+\bm{p}}{|\bm{p}'+\bm{p}|},\quad
\hat{\bm{m}}=\frac{\bm{p}'-\bm{p}}{|\bm{p}'-\bm{p}|},\quad
\hat{\bm{n}}=\frac{\bm{p}\times\bm{p}'}{|\bm{p}\times\bm{p}'|}
\end{equation}
for $\bm{p}$ and $\bm{p}'$ as the initial and final momenta of the
proton. The first term is the central contribution, the next three
terms are double-spin-flip terms and the final term is a
single-spin-flip term. We can determine the sensitivity of the
$(e,e'p)$ observables to these terms by determining the Saclay
amplitudes $a_s$, $b_s$, $c_s$, $d_s$ and $e_s$ from the helicity
matrix elements as described in the Appendix, setting some of the
amplitudes to zero and then transforming the result to give new
invariant amplitudes for the Fermi form. A common approximation is
to use only the central part of the amplitude generated from a
prescription involving the total cross section.

\begin{figure}
\centerline{\includegraphics[height=2.5in]{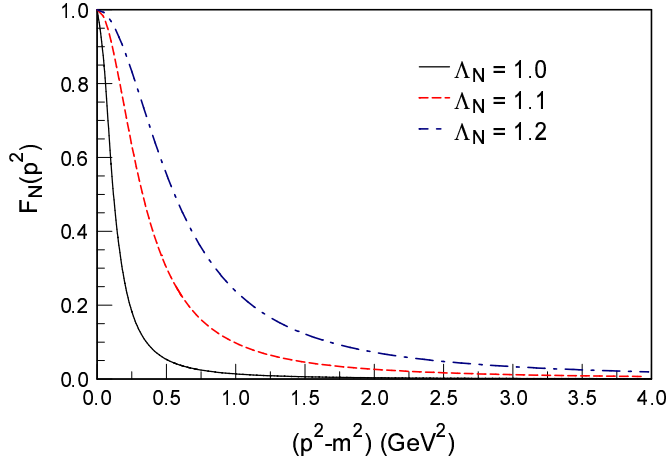}}
\caption{Off-shell nucleon form factor for $\Lambda_N=$ 1.0, 1.1 and 1.2 GeV.  }\label{offshell}
\end{figure}

The contribution to the current matrix element given by
(\ref{J_FSIb}), represented by Fig. \ref{Jabc}b, involves a
principal value integral over off-mass-shell momenta for one leg of
the scattering amplitude. The proton propagator for this leg
contains only the positive energy contribution.  Determination of
the off shell behavior of the scattering amplitude requires a
dynamical model of the amplitude.  Such a model is not currently
available to us in the range of energies required for the
experiments being performed at Jefferson Lab. In order to estimate
the possible effects of this contribution to the current matrix
elements, we use a simple prescription for the off-shell behavior of
the amplitude. Although additional invariants are possible when the
nucleon is allowed to go off shell, we keep only the forms in
(\ref{Fermi}).  The center-of-momentum angle is calculated using

\begin{equation}
\displaystyle{\cos\theta=\frac{t-u}{\sqrt{s-4m^2}\sqrt{\frac{(4m^2-t-u)^2}{s}-4m^2}}}\label{thetacm}
\end{equation}
The invariants are then replaced by
\begin{equation}
\mathcal{F}_i(s,t)\rightarrow\mathcal{F}_i(s,t,u)F_N(s+t+u-3m^2)
\end{equation}
where
\begin{equation}
F_N(p^2)=\frac{(\Lambda_N^2-m^2)^2}{(p^2-m^2)^2+(\Lambda_N^2-m^2)^2}\label{Nff}
\end{equation}
and the $\mathcal{F}_i(s,t,u)$ are obtained from interpolation of
the on-shell invariant functions with the center-of-momentum angle
obtained from (\ref{thetacm}). The form factor (\ref{Nff}) was used as a cutoff in
calculating the Gross vertex function used in this paper with $\Lambda_N=1.675$ GeV. However, there should be an intrinsic fall off of the scattering amplitude due to the dynamics of the scattering which would be expected to be faster than provided by this cutoff mass.  Figure \ref{offshell} shows the off-shell form factor for cutoff masses of $\Lambda_N=$ 1.0, 1.1 and 1.2 GeV. In the absence of a dynamical model of the scattering amplitudes, the effect of possible off-shell contributions on various observables can be reasonably estimated by using cutoff masses in this range.

The contribution to the current matrix elements from (\ref{J_FSIc}),
represented by Fig. \ref{Jabc}c, contains the effect of negative
energy propagation of the off-shell leg of the scattering
amplitude. Since the denominator of this part of the propagator will
be large compared with that of the positive energy part at large
momentum transfers, we neglect this contribution for the present.
This is the only approximation involved in our calculation.  
Note, however, that by dropping this contribution the current matrix
elements are no longer covariant. We have chosen to calculate the
matrix elements in the laboratory frame.

\subsection{Differential Cross Section}

The $D(e,e'p)$ cross section for unpolarized deuterons and protons in the lab frame can be written as
\cite{raskintwd,dmtrgross}

\begin{eqnarray}
\left ( \frac{ d \sigma^5}{d \epsilon' d \Omega_e d \Omega_p} \right
)_h  & = & \frac{m_p \, m_n \, p_p}{8 \pi^3 \, M_d} \,
\sigma_{Mott} \,
f_{rec}^{-1} \, \nonumber \\
& & \Big[ \left ( v_L R_L +   v_T R_T
 + v_{TT} R_{TT}\cos 2\phi_p + v_{LT} R_{LT}\cos\phi_p \right )
  \nonumber \\
& & +  h   v_{LT'} R_{LT'}\sin\phi_p
\Big] \, , \label{xsdef}
\end{eqnarray}
where $M_d$, $m_p$ and $m_n$  are the masses of the deuteron, proton and neutron,
 $p_p=p_1$ and $\Omega_p$
are the momentum and solid angle of the ejected proton, $\epsilon'$ is the
energy of the detected electron and $\Omega_e$ is its solid angle.
The helicity of the electron is denoted by $h$. The Mott cross
section is
\begin{equation}
\sigma_{Mott} = \left ( \frac{ \alpha \cos(\theta_e/2)} {2
\varepsilon \sin ^2(\theta_e/2)} \right )^2
\end{equation}
and the recoil factor is given by
\begin{equation}
f_{rec} = \left| 1+ \frac{\omega p_p - E_p q \cos \theta_p} {M_d \, p_p}
\right| \, . \label{defrecoil}
\end{equation}
The kinematic coefficients $v_K$ are
\begin{eqnarray}
v_L&=&\frac{Q^4}{q^4}\\
v_T&=&\frac{Q^2}{2q^2}+\tan^2\frac{\theta_e}{2}\\
v_{TT}&=&-\frac{Q^2}{2q^2}\\
v_{LT}&=&-\frac{Q^2}{\sqrt{2}q^2}\sqrt{\frac{Q^2}{q^2}+\tan^2\frac{\theta_e}{2}}\\
v_{LT'}&=&-\frac{Q^2}{\sqrt{2}q^2}\tan\frac{\theta_e}{2}
\end{eqnarray}
If the response tensor is defined as
\begin{equation}
W^{\mu\nu}=\frac{1}{3}\sum_{s_1,s_2,\lambda_d}\left<\bm{p}_1s_1;\bm{p}_2s_2\right|J^\mu\left|\bm{P}\lambda_d\right>^* \left<\bm{p}_1s_1;\bm{p}_2s_2\right|J^\nu\left|\bm{P}\lambda_d\right>
\end{equation}
the response functions $R_K$ are defined by
\begin{eqnarray}
R_L & \equiv & W^{00} \nonumber \\
R_T & \equiv & W^{11}+W^{22}=w_{1,1}+w_{-1,-1}  \nonumber  \\
R_{TT}\cos 2\phi_p & \equiv &  W^{22}-W^{11}=2\Re(w_{1,-1}) \nonumber  \\
R_{LT}\cos\phi_p & \equiv & 2 \sqrt{2}\, \Re(W^{01})=-2\Re(w^0_1-w^0_{-1})   \nonumber \\
R_{LT'}\sin\phi_p & \equiv & - 2 \sqrt{2}\, \Im(W^{01})=-2\Re(w^0_1+w^0_{-1}) \, , \label{defresp}
\end{eqnarray}
where
\begin{equation}
w_{\lambda'_\gamma,\lambda_\gamma}=\frac{1}{3}\sum_{s_1,s_2,\lambda_d}
\left<\bm{p}_1s_1;\bm{p}_2s_2\right|J_{\lambda'_\gamma}\left|\bm{P}\lambda_d\right>^* \left<\bm{p}_1s_1;\bm{p}_2s_2\right|J_{\lambda_\gamma}\left|\bm{P}\lambda_d\right>
\end{equation}
and
\begin{equation}
w^0_{\lambda_\gamma}=\frac{1}{3}\sum_{s_1,s_2,\lambda_d}
\left<\bm{p}_1s_1;\bm{p}_2s_2\right|J^0\left|\bm{P}\lambda_d\right>^* \left<\bm{p}_1s_1;\bm{p}_2s_2\right|J_{\lambda_\gamma}\left|\bm{P}\lambda_d\right>
\end{equation}
with
\begin{equation}
J_{\pm 1}=\mp\frac{1}{\sqrt{2}}(J^1\pm J^2)
\end{equation}

For our
calculations, we have chosen the following kinematic conditions: the
z-axis is parallel to $\bm{ q}$, and the missing momentum is defined
as $\bm{p}_m \equiv \bm{q} - \bm{p}_1=\bm{p}_2=\bm{p}_n$.

\subsection{Asymmetries}

The representation of the cross section in terms of response functions is due to the mixed polarization of the virtual photon which varies with the electron kinematics and polarization. The transverse-transverse response function is the result of interference between the $\lambda_\gamma=\pm 1$ helicity states while the longitudinal-transverse response functions $R_{LT}$ and $R_{LT'}$ are the result of interference between the deuteron charge and two linear combinations of the $\lambda_\gamma=\pm 1$ helicity states. As an alternative to a complete separation of the cross section into response functions, the interference response functions can be accessed through linear combinations of differential cross sections to produce three interference asymmetries defined as
\begin{equation}
A_{TT}=\frac{v_{TT}R_{TT}}{v_L R_L+v_T R_T}\label{defatt}
\end{equation}

\begin{equation}
A_{LT}= \frac{\sigma_{0}(0^\circ) -
\sigma_{0}(180^\circ)} {\sigma_{0}(0^\circ) +
\sigma_{0}(180^\circ)  } = \frac{v_{LT}R_{LT}}{v_L R_L+v_T R_T+v_{TT}R_{TT}}\label{defalt}
\end{equation}
and
\begin{equation}
\displaystyle{ A_{LT'} = \frac{\sigma_{+1}(90^\circ) -
\sigma_{-1}(90^\circ)} {\sigma_{+1}(90^\circ) +
\sigma_{-1}(90^\circ)  }
 =  \frac{v_{LT'} R_{LT'}}{v_L R_L + v_T R_T -
v_{TT} R_{TT} }} \label{defatlp}  \,,
\end{equation}
where, for conciseness,
\begin{equation}
\sigma_{h}(\phi_p)\equiv \left ( \frac{ d \sigma^5}{d \epsilon' d \Omega_e d \Omega_p} \right
)_h
\end{equation}
Note that while $A_{LT}$ can be obtained by measuring protons in the electron scattering plane symmetrically about the direction of the three-momentum transfer, the asymmetries $A_{TT}$ and $A_{LT'}$ require measurements to be made out of the scattering plane.
The asymmetry $A_{LT'}$ is defined as an electron single spin asymmetry and can, therefore, be easily obtained by flipping the beam helicity. While $R_L$ and $R_T$ are independent of photon-helicity-dependent phases, the interference response functions are not.  As a result, the interference response functions can be very sensitive to phase differences generated by non-nucleonic currents and final state interactions.  This is particularly true of $R_{LT'}$ which can be shown to be zero in the PWIA. The interference
response function, $R_{LT}$, is very sensitive to the relativity included in
the current operator, due to the various interference contributions
from the charge and transverse current operators \cite{relcur}.

The observable $A_{LT'}$ has recently been measured in Jefferson Lab's
Hall B \cite{jerrygexp}. Due to the large solid angle coverage in
Hall B, and the averaging over $\phi_p$, the transverse-transverse
interference response that is multiplied with a factor of $\cos(2\phi_p)$ in the cross section, drops out of the measured asymmetry
$A_{LT'}$ \cite{jerrygprivcom}. Therefore, from now on in this
paper, we calculate
\begin{equation}
\displaystyle{ A_{LT'}^{Hall \, B}   = \frac{ v_{LT'} R_{LT'}}{v_L
R_L + v_T R_T }} \label{defatlphallb} \,.
\end{equation}
The difference between the asymmetry calculated with and without
$R_{TT}$ is very small in practice, due to the small size of the
transverse-transverse response.

\section{Results}

In this section, we discuss our numerical results for several
observables. We will investigate the effect of the final state
interactions (FSIs), and in particular, we will point out the
contributions of spin-dependent FSIs, both the single-spin-flip
contributions and the double-spin-flip contributions. We will also
discuss the relative importance of on-shell and off-shell
contributions to FSI.

\subsection{Differential Cross Sections}

\subsubsection{Momentum Distributions}

\begin{figure}[ht]
\includegraphics[width=20pc,angle=270]{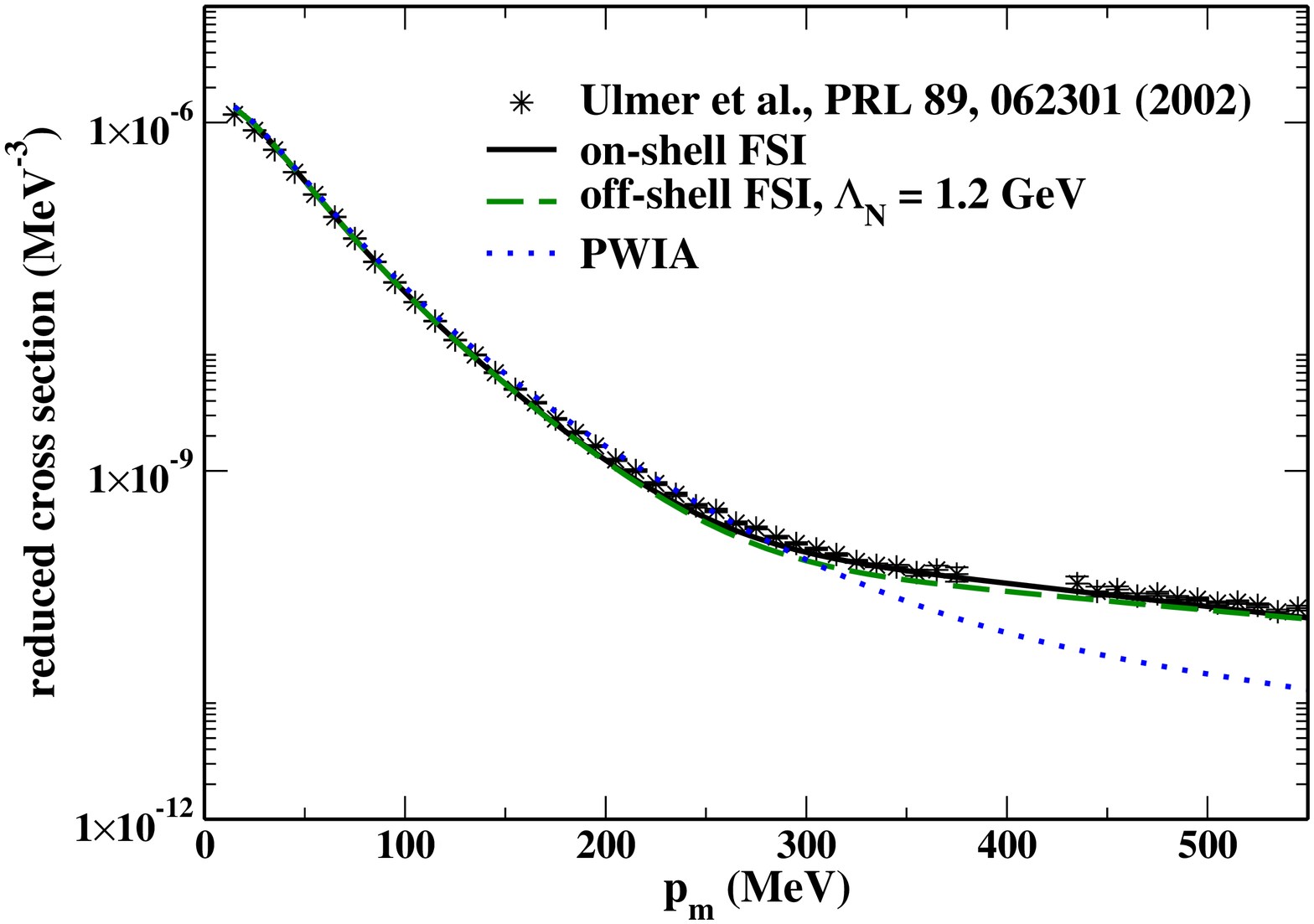}
\includegraphics[width=20pc,angle=270]{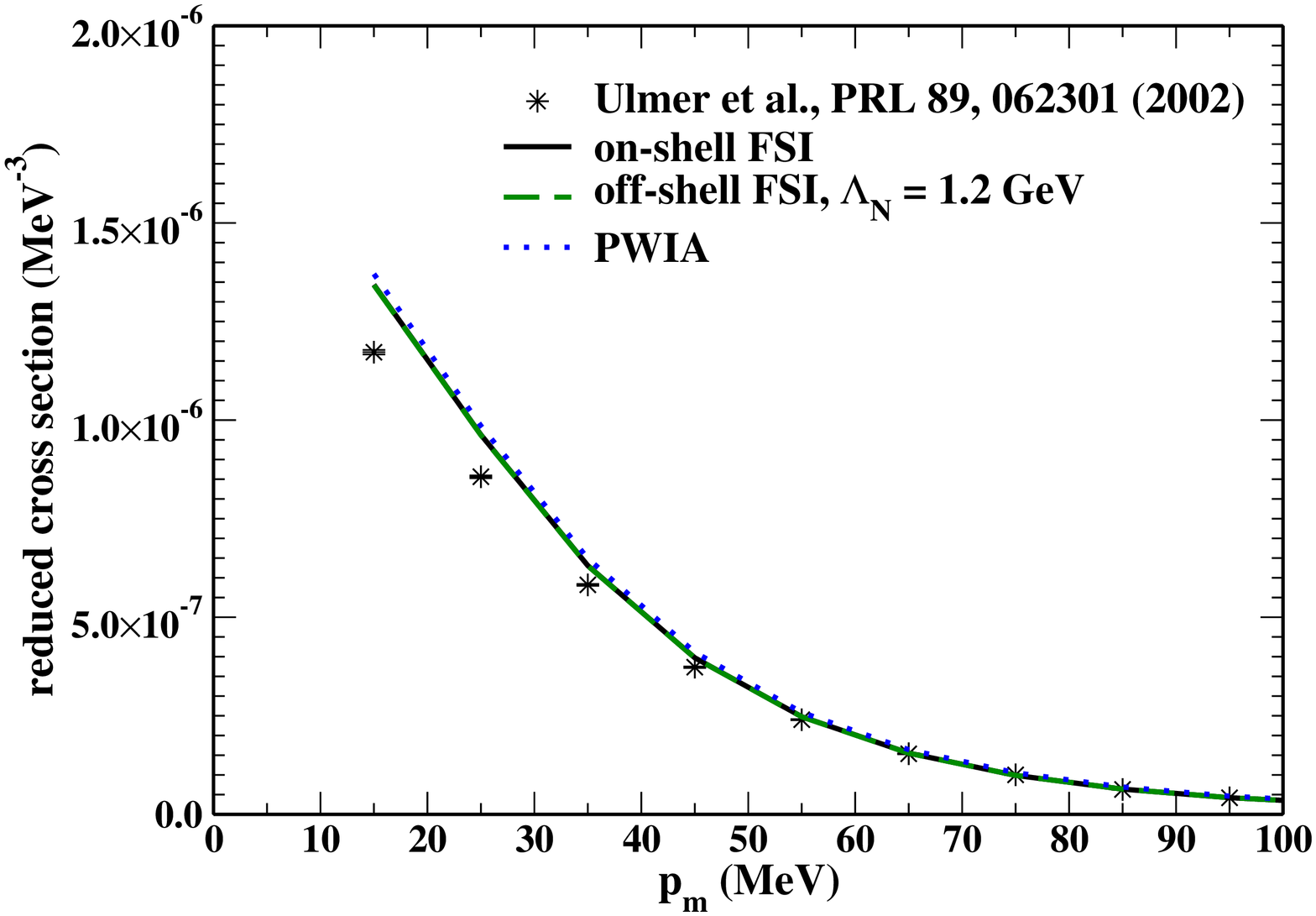}
\caption{The reduced cross section for a beam energy of $3.1095
$ GeV, $Q^2 = 0.665~{\rm GeV}^2$, $x_{Bj} = 0.964$, and $\phi_p = 180^\circ$.
The data are from \protect{\cite{paulprl}}.
 } \label{figpaul}
\end{figure}

The most natural observable to investigate is the differential cross
section. In Fig. \ref{figpaul} our calculations are compared to the published
$D(e,e'p)$ data from Jefferson Lab's Hall A. The data from Ulmer et
al. \cite{paulprl} are shown together with our PWIA, on-shell FSI,
and full FSI curves as a function of the missing momentum $p_m$.
For all of the calculations in the paper we use the MMD \cite{mmd}
proton electromagnetic form factors. The data are presented
as a reduced cross section, which is defined as
\begin{equation}
\sigma_{reduced} = \frac{d^5 \sigma}{d \Omega_{e'} d \Omega_N d
E_{e'}} \, \, \frac{M_df_{rec}}{\sigma_{ep} m_p m_n p_p}  \,.
 \label{defredsigma}
\end{equation}

At lower missing momenta, the effects of FSIs are small. For missing
momenta larger than $p_m \approx 110$ MeV, the FSI effects become
visible. First, the PWIA curve is slightly above the FSI results,
then, at $p_m \approx 300$ MeV, the FSI curves become larger than
the PWIA contribution.  The agreement with the data is quite nice
overall. The off-shell FSI is small, and leads to final results a
little below the data at larger missing momenta. This is a sensible
result, as we do expect meson exchange currents (MECs) to play a
role at the relatively low $Q^2$ at which these data were taken.
Indeed, in \cite{paulprl}, the calculation by Arenhoevel
\cite{arenhoevel} agreed with the data at large missing momenta after MEC contributions were
included; the FSI-only calculation was a bit below the data.

We also have added a panel with a linear plot of just the low
missing momentum data. The full FSI curve and the on-shell FSI curve
coincide in this region. It has been observed in several previous
$D(e,e'p)$ measurements that at very low missing momenta, the
calculations are somewhat above the data. This is quite puzzling as
at these low missing momenta, effects like FSIs, MECS etc are
supposed to be small and well under control. For a nice compilation
on this topic, see \cite{boeglintrento2005}. In \cite{paulprl}, Fig.
1 shows the deviation of the reduced cross section data and the
calculations. Here, we observe the same type of deviation at very
low $p_m \leq 0.35~MeV$. The largest discrepancy appears at $p_m =
15~MeV$, where our calculation overpredicts the data by $15\%$.
Overall, comparing with the previous results, our low missing
momentum results seem to be an improvement, even though the
discrepancy has not been fully removed. One main difference between
the calculation presented in this article and the calculations
in\cite{paulprl} is the fully relativistic approach we take here.

\begin{figure}[ht]
\includegraphics[height=4in,angle=270]{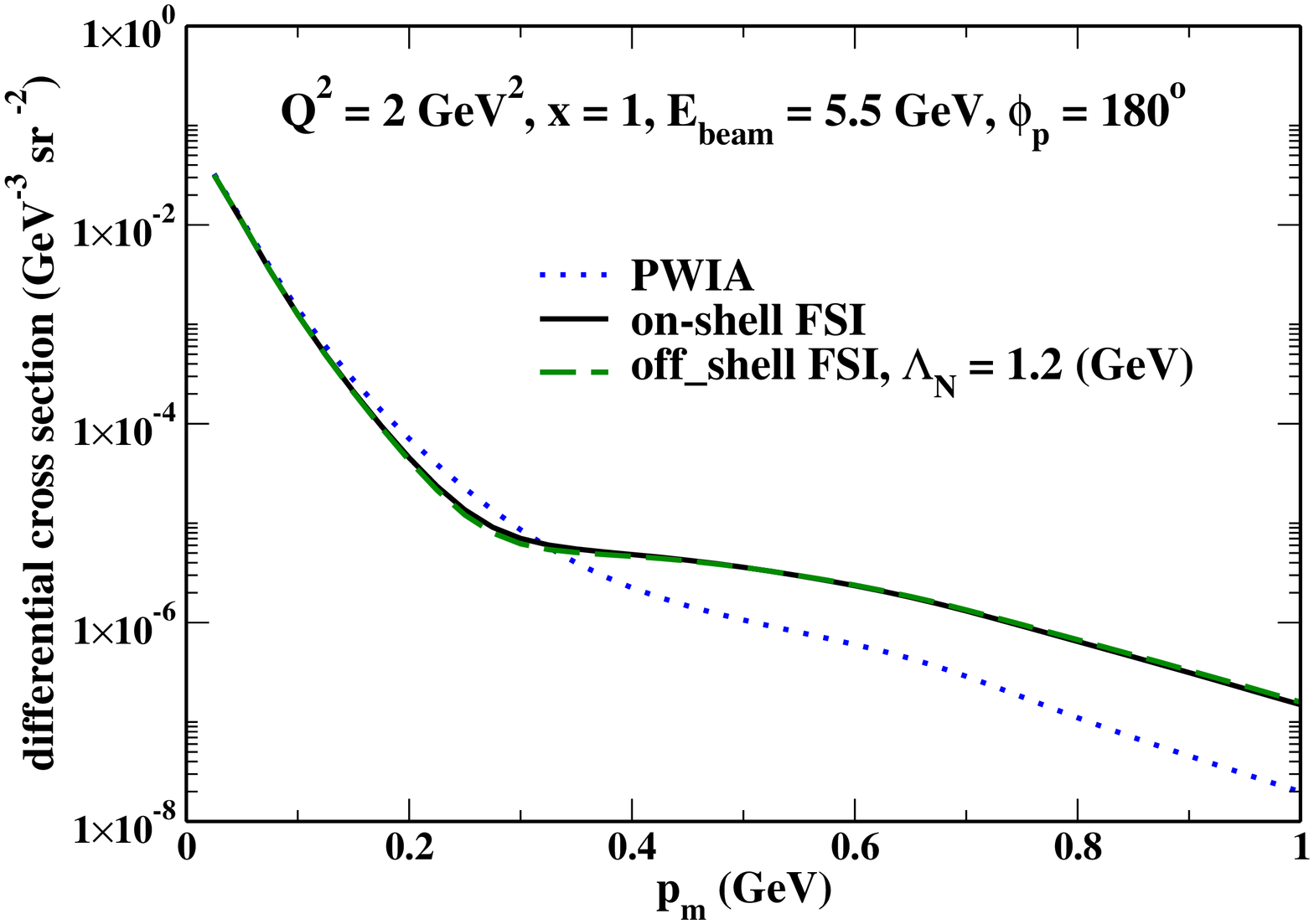}
\includegraphics[height=4in,angle=270]{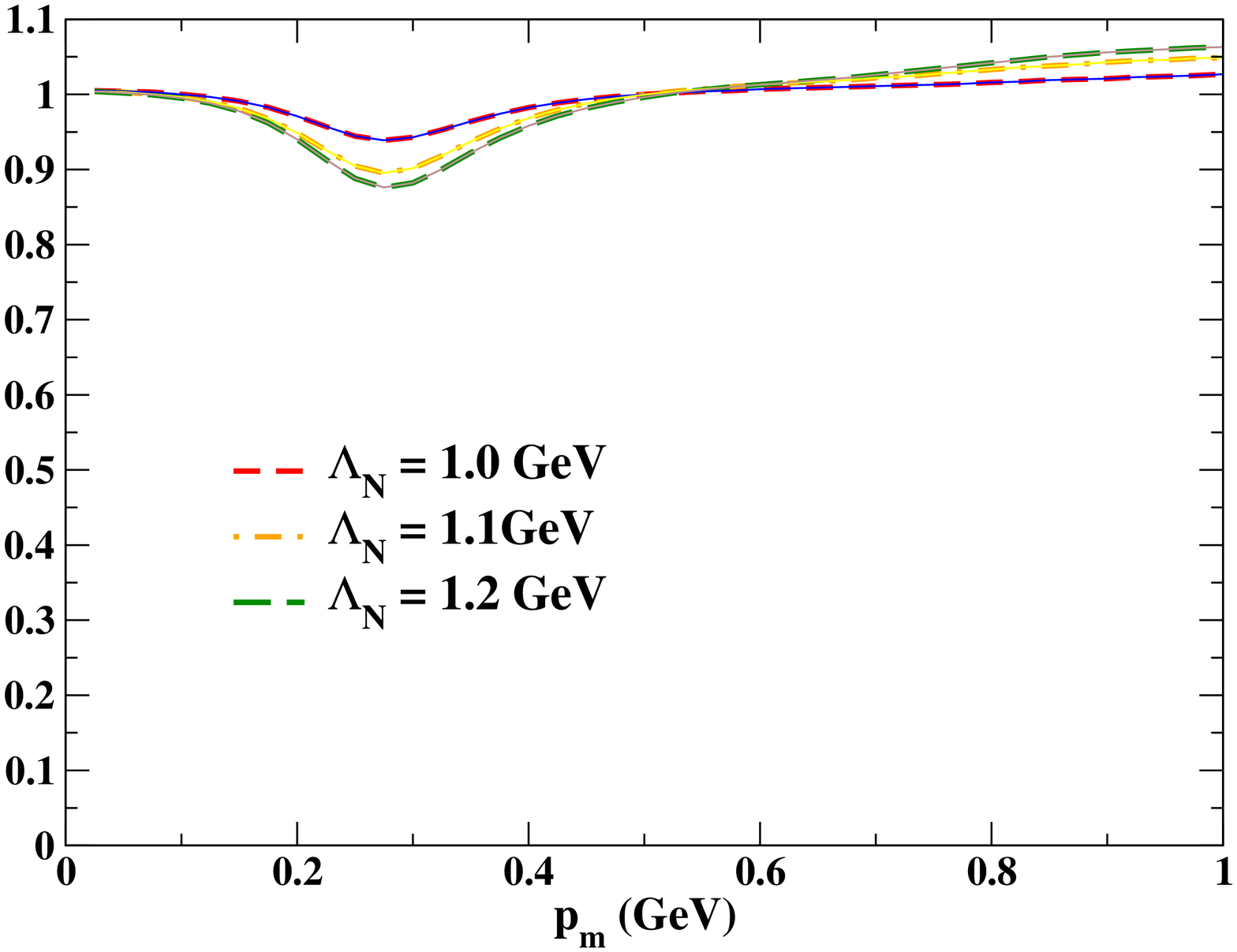}
\caption{Top panel: the differential cross section for a beam energy of $5.5
{\rm GeV }$, $Q^2 = 2~{\rm GeV}^2$, $x_{Bj} = 1$, and $\phi_p = 180^\circ$ is shown
in PWIA (dotted curve), with on-shell FSI (solid curve), and with
on-shell and off-shell FSIs (dashed curve), as a function of the
missing momentum. Bottom panel: the ratio of the off-shell calculations
with varying cut-off $\Lambda_N = $ 1 GeV (short-dashed),
 1.1 GeV (dash-dotted), and 1.2 GeV (long-dashed) to the
on-shell FSI calculation, in the same kinematics as the top panel.}
\label{figcsmomdis}
\end{figure}

In Figure \ref{figcsmomdis}, the upper panel shows the cross section as a
function of the missing momentum, with a beam energy of $5.5$ GeV,
$Q^2 = 2~{\rm GeV}^2$, $x_{Bj} = 1$, and $\phi_p = 180^\circ$. The choice of
Bjorken-x implies roughly quasi-free kinematics. The azimuthal angle
of the detected proton has been chosen to maximize the cross
section. The PWIA and on-shell FSI curves are almost identical at
very low missing momenta, up to $0.05$ GeV. Then, the FSI curve
reduces the cross section compared to the PWIA result, roughly from
$p_m = 0.05$ GeV to $p_m = 0.33$ GeV. For larger missing momenta,
there is a marked increase in the differential cross section when
FSI is included. These results are quite typical and have been seen
in other calculations \cite{sofsi,misak,laget,ciofi}. The
differential cross section decreases by several orders of magnitude with
increasing missing momentum, and a small reduction due to FSI at
lower missing momenta and larger differential cross section can lead
to a very large increase at larger missing momenta and smaller
differential cross sections. The inclusion of the off-shell FSI
contributions leads to a slight reduction of the cross section for
medium missing momenta.

The lower panel shows the ratios of the off-shell FSI calculations with $\Lambda_N=$ 1.0, 1.1 and 1.2 GeV to the on-shell FSI. The off-shell effects are small for small $p_m$, but become increasingly large as $p_m$ increases.  These effects are not particularly sensitive to the cutoffs chosen here, but must be quite sensitive to lower cutoff masses.

\begin{figure}[ht]
\includegraphics[width=20pc,angle=270]{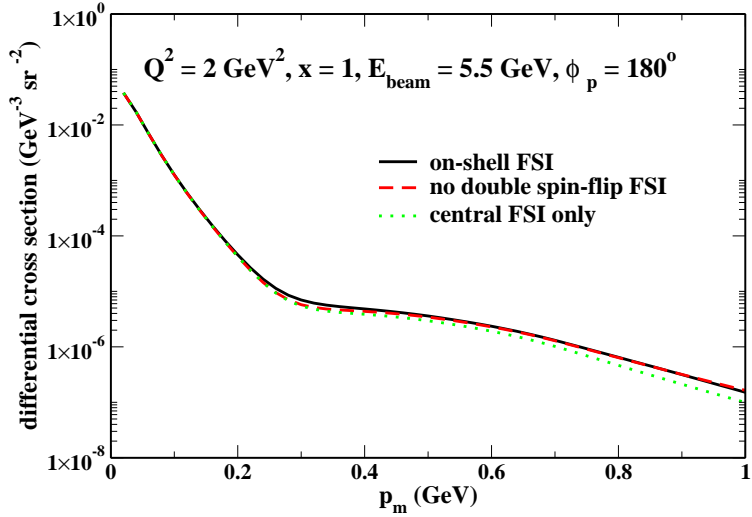}
\caption{The differential cross section for a beam energy of $5.5~
{\rm GeV}$, $Q^2 = 2 ~{\rm GeV}^2$, $x_{Bj} = 1$, and $\phi_p = 180^\circ$ is shown
calculated with on-shell FSI, as a function of the missing momentum.
The solid line shows the result calculated with the full $NN$
scattering amplitude, the dashed line shows the result without the
double-spin-flip terms of the $NN$ scattering amplitudes, and the
dotted line shows the result with the central $NN$ amplitude only. }
\label{figcsmdspin}
\end{figure}

In Fig. \ref{figcsmdspin}, we investigate the role played by the
various parts of the proton-neutron scattering amplitude contributing to
(\ref{Fermi}). The dashed line shows the results without the
three double-spin-flip terms in the amplitude. One can see that for
missing momenta from $0.2$ GeV to $0.45$ GeV, the double-spin-flip
contribution is quite relevant. Its omission in this region leads to
a significantly smaller cross section. The single-spin-flip - or
spin-orbit - part of the proton-neutron scattering amplitude becomes
relevant only at higher missing momenta than the double-spin-flip
terms. From roughly $p_m = 0.3$ GeV on, omitting the spin-orbit
contribution leads to a decrease in the differential cross section.

This clearly shows that while it is possible to parameterize the
central part of the $NN$ scattering amplitude, and to reproduce the
$NN$ cross section data this way, this type of parametrization
effectively includes some physics that stems from the spin-dependent
parts of the $NN$ amplitude.  Here, it is interesting to see the
influence of the spin-dependent parts of the $NN$ amplitude on the
unpolarized cross section. While the logarithmic scale necessary for
the momentum distribution conceals the effects, it is important to
note that the relative importance of the spin-dependent FSI
contributions changes with missing momentum. We will return to the
effects of spin-dependent FSI with Fig. \ref{figcsadspin}.

\begin{figure}[ht]
\includegraphics[height=4in,angle=270]{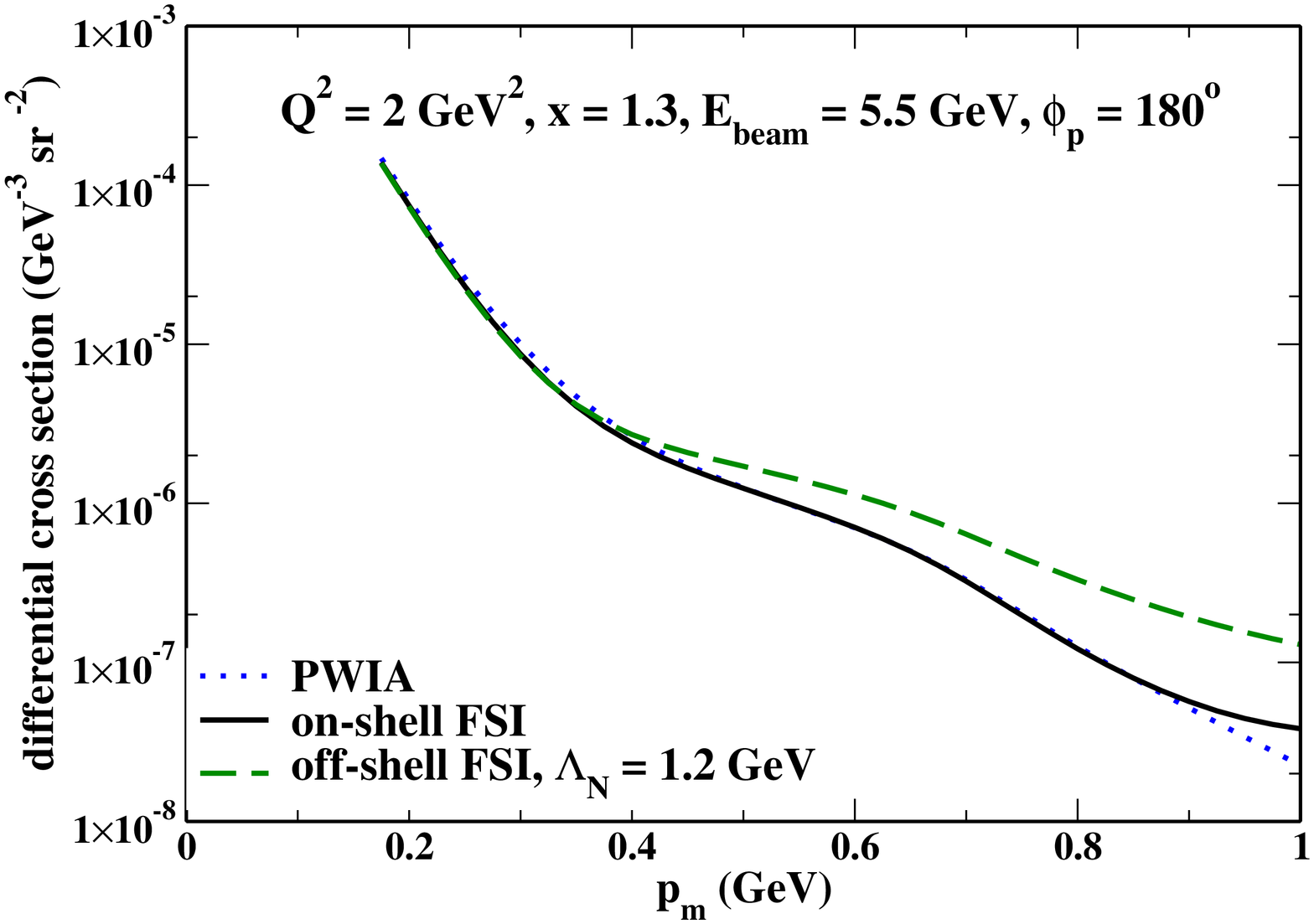}
\includegraphics[height=4in,angle=270]{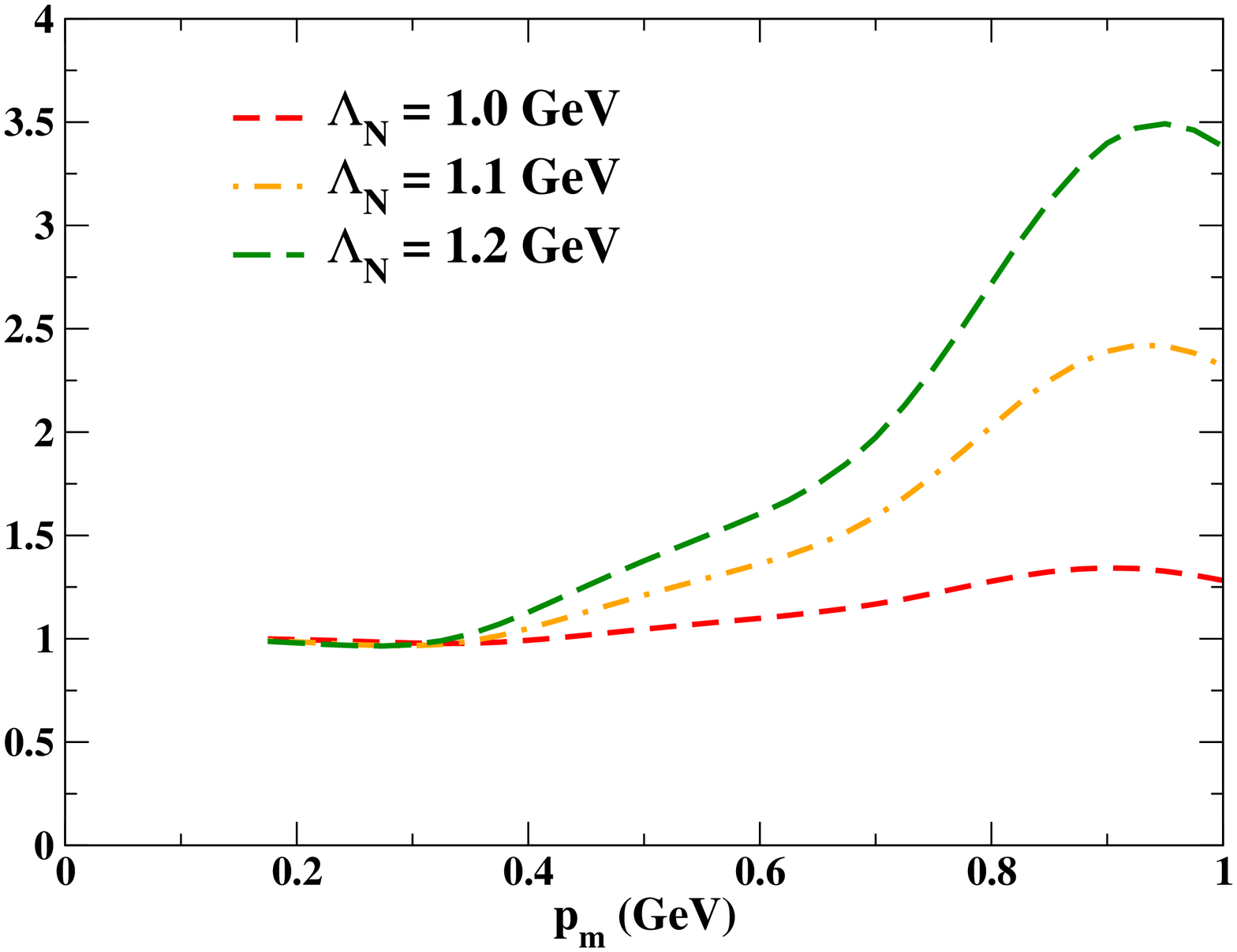}
\caption{Top panel: the differential cross section for a beam energy of $5.5
$ GeV, $Q^2 = 2~{\rm GeV}^2$, $x_{Bj} = 1.3$, and $\phi_p = 180^\circ$ is shown
in PWIA (dotted curve), with on-shell FSI (solid curve), and with
on-shell and off-shell FSIs (dashed curve), as a function of the
missing momentum. Bottom panel: the ratio of the off-shell calculations
with varying cut-off $\Lambda_N = $ 1 GeV (short-dashed),
 1.1 GeV (dash-dotted), and 1.2 GeV (long-dashed) to the
on-shell FSI calculation, in the same kinematics as the top panel.} \label{figcsmdx1p3}
\end{figure}

In Fig. \ref{figcsmdx1p3}, we display our results for the same
four-momentum transfer, but higher $x_{Bj} = 1.3$. At this value of
$x_{Bj}$, and larger values,  strong short-range $pn$ correlations
have been reported by inclusive electron scattering experiments on
deuterium \cite{egiyansrc,piasetzky}. The deviation of the on-shell
FSI from the PWIA is small for medium missing momenta, and seems to
disappear altogether for missing momenta between 0.5 GeV and 0.6 GeV.
However, the
off-shell contribution to the FSI gains in relevance for larger
missing momenta, and leads to a significant increase over the PWIA
results at $p_m > 0.4$ GeV. The lower panel again shows the ratio of off-shell FSI to the on-shell result for three values of the off-shell cutoff. The importance of the off-shell FSI here
is larger than for $x_{Bj} = 1$ (as discussed above when considering
Fig. \ref{figcsmomdis} and the sensitivity to the value of the cutoff is much greater. This is the expected behavior, as the
deviation from $x_{Bj} = 1$ corresponds to a deviation from the
quasi-elastic kinematics, and stresses the off-shell region more. A
recent new proposal \cite{wernernewprop} suggests a measurement of
the cross section at somewhat larger $Q^2 = 3.5~{\rm GeV}^2$, but the same
value of $x_{Bj} = 1.3$ and a beam energy of $E_{beam} = 5.25$ GeV.
The results of our calculation for these kinematics are similar to
what is displayed above in Fig. \ref{figcsmdx1p3}. The on-shell
FSI in this case deviates a bit more from the PWIA result than for
the kinematics displayed here. The off-shell contribution is just as
significant in the proposed kinematics.

\subsubsection{Angular Distributions}

\begin{figure}[ht]
\includegraphics[width=20pc,angle=270]{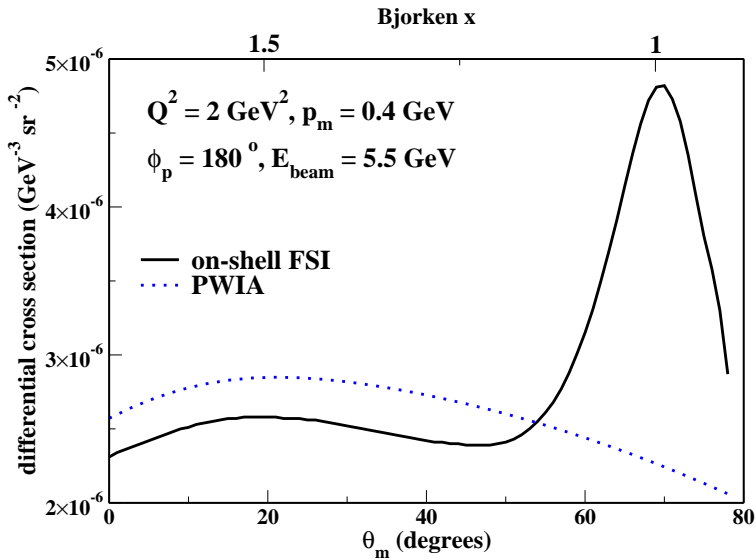}
\caption{The differential cross section for a beam energy of $5.5
$ GeV, $Q^2 = 2~{\rm GeV}^2$, $p_{m} = 0.4$ GeV, and $\phi_p = 180^\circ$ is
shown in PWIA (dotted curve), with on-shell FSI (solid curve) FSI (dashed curve), as a function of
the angle of the missing momentum. The top axis shows the
corresponding values of Bjorken-x. } \label{figcsangdis}
\end{figure}

In Figure \ref{figcsangdis}, we show the cross section as a
function of the angle $\theta_m$ of the missing momentum, with fixed
$Q^2 = 2~{\rm GeV}^2$ and $p_m = 0.4$ GeV. The beam energy and azimuthal
angle of the proton are the same as for the momentum distribution
graphs. The angular distribution of the cross section shows much
less variation in the magnitude, and therefore can be shown on a
linear plot, allowing for a better look at the relevance of various
parts of the cross section. While the region beyond $\theta_m =
80^\circ$ is kinematically accessible to experiment, a calculation in
this region requires the knowledge of the $N N$ scattering amplitude
above lab kinetic energies of $1.3$ GeV in the $pn$ system, which
are not available from SAID. We therefore stay below $80^\circ$. As is
obvious from the plot, the most interesting features of the
calculation are located below this angle: while the PWIA results are
gently sloping upwards and then downwards for angles larger than $20^{\circ}$, the
FSI results initially follow this behavior, but then show a
pronounced peak at around $\theta_m = 70^\circ$. This value corresponds
to $x_{Bj} = 1$, i.e. quasi-free kinematics for the knocked-out
nucleon. For the lower angles, the FSI simply leads to a reduction
in the cross section, but the shape is unchanged. For larger angles,
around $x_{Bj} = 1$, the diffractive nature of the FSI leads to a
redistribution of strength from smaller missing momenta, causing a
large peak.

If we consider the ratio of FSI to PWIA cross section, as is
sometimes done when comparing various methods of calculation
\cite{misak,boeglintrento2005}, this ratio peaks at $70^\circ$, too. Our
calculation clearly shows the same shift from a peak at $90^\circ$, as
seen in Glauber theory calculations \cite{sabineglauber}, to a lower
angle, as seen in the Generalized Eikonal Approximation (GEA)
\cite{misak} and the diagrammatic approach of Laget \cite{laget}.

\begin{figure}[ht]
\includegraphics[width=20pc,angle=270]{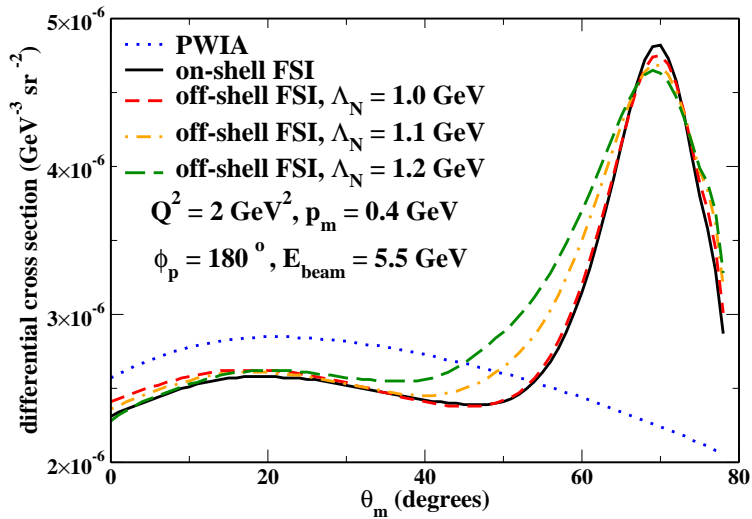}
\caption{The differential cross section for a beam energy of $5.5
$ GeV, $Q^2 = 2 ~{\rm GeV}^2$, $p_{m} = 0.4 $ GeV, and $\phi_p = 180^\circ$ is
shown with on-shell FSI (solid curve), and with on-shell and
off-shell FSIs at various cut-offs, as a function of the angle of
the missing momentum. } \label{figcsadoffshell}
\end{figure}

For the angular distribution, the influence of the off-shell FSI cut-off
is clearly visible and can be studied easily. One does not expect a large contribution from far off-shell
nucleons. The cut-off we use here serves to impose that constraint.
For the kinematics displayed in Fig. \ref{figcsadoffshell}, we
investigate the effects of various cut-off values. The cut-off at
$1$ GeV leads to a very small increase at small angles and a very
small decrease at large angles, but the overall result is hardly
different from the on-shell FSI only result. The primary effect for larger cutoff masses is to fill in the minimum in the on-shell result from $30^\circ$ to $65^\circ$.

\begin{figure}[ht]
\includegraphics[width=20pc,angle=270]{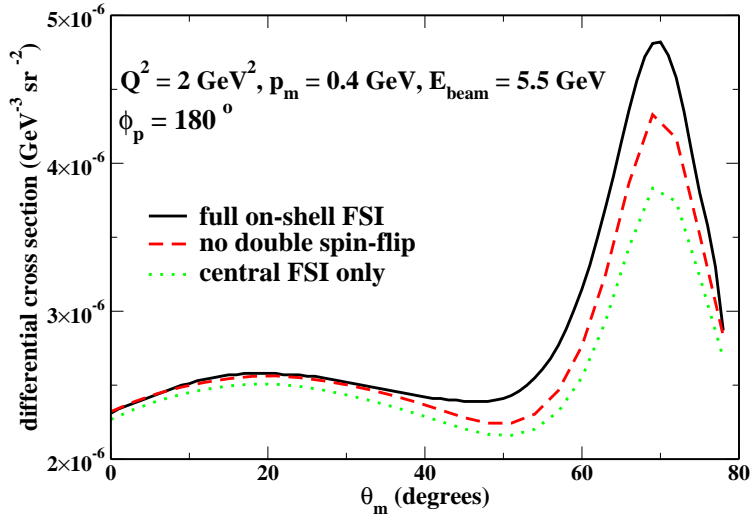}
\caption{The differential cross section for a beam energy of $5.5
$ GeV, $Q^2 = 2~{\rm GeV}^2$, $p_{m} = 0.4$ GeV, and $\phi_p = 180^\circ$ is
shown calculated with on-shell FSI, as a function of the angle of
the missing momentum. The solid line shows the result calculated
with the full $NN$ scattering amplitude, the dashed line shows the
result without the double-spin-flip terms of the $NN$ scattering
amplitudes, and the dotted line shows the result with the central
$NN$ amplitude only. } \label{figcsadspin}
\end{figure}

In Fig. \ref{figcsadspin}, we show the effects of the various
spin-dependent parts of the $ p n$ scattering amplitude. It is very
interesting to observe that in the peak region, the contribution of
the spin-dependent FSIs (both single and double-spin-flip) amounts
to about one quarter of the cross section. This is certainly a
rather significant contribution. The contribution from the single-spin-flip term and the three double-spin-flip terms is about equal
in the peak region. The figure shows that the double-spin-flip
contribution to the cross section at small angles is almost
negligible. It becomes noticeable at $\theta_m \approx 40^\circ$, and
then leads to a sizable increase of the differential cross section
in the peak region. The omission of the single-spin-flip
contribution leads to a noticeable reduction in the cross section
for all angles. The effect is most pronounced in the peak region and
in the very shallow dip just before the peak region.

\begin{figure}[ht]
\includegraphics[width=20pc,angle=270]{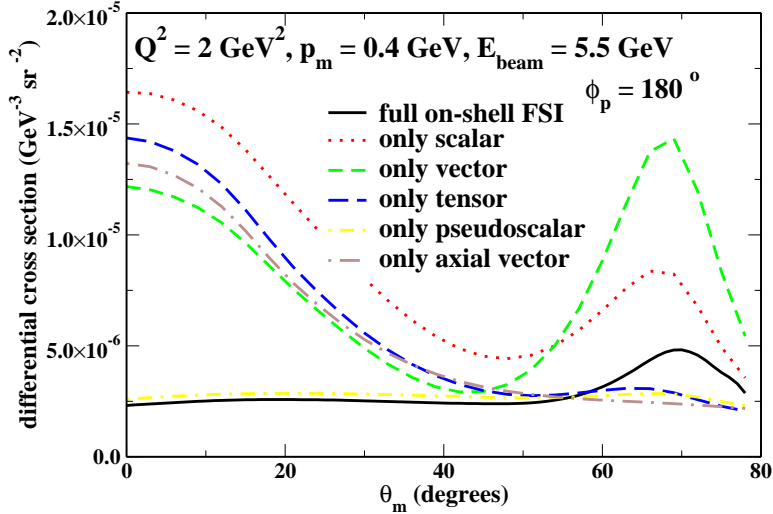}
\caption{The differential cross section for a beam energy of $5.5
$ GeV, $Q^2 = 2~{\rm GeV}^2$, $p_{m} = 0.4$ GeV, and $\phi_p = 180^\circ$ is
shown with on-shell FSI  as a function of the angle of the missing
momentum. The solid curve shows the contribution of the full $NN$
scattering amplitude, the other curves show the results for just one
invariant term in the $NN$ scattering amplitude. }
\label{figcsadinv1}
\end{figure}

To this point, we have considered the different contributions of the
spin-dependent parts of the amplitude, i.e. of the $NN$ amplitude
split up following the Saclay convention (\ref{saclaynndef}). Using the Saclay formalism with its
classification according to the spin-dependence is quite useful, as
it allows one to understand the new contributions from different
parts of the current operator when adding the single-spin-flip term
and the double-spin-flip terms, see \cite{sofsi}. Even though we did
not rewrite our current operator to distinguish e.g. between
magnetization current and convection current, seeing the $NN$
scattering amplitude in terms of its spin-dependence is a very
natural view point, and allows for a certain intuitive understanding
of the numerical results.

It is also interesting to investigate the $NN$ amplitude in terms of
the five Fermi invariants, which are so practical for actual calculations.
>From (\ref{HetoInv},\ref{HetoSa}), it is clear that every
invariant amplitude contains several different, spin-dependent
pieces. In Fig. \ref{figcsadinv1}, we show the results of the
calculation with on-shell FSI if only one of the five invariants is
included. For comparison, the solid lines depicts the result
obtained with the full $NN$ amplitude. The result obtained with just
the pseudoscalar part is slightly above the full result for smaller
angles, and continues smooth and almost straight towards larger
angles. It does not exhibit a peak structure at large angles. The
tensor and axial vector contributions are fairly close to the
pseudoscalar contribution at large angles, and neither exhibits a
peak at large angles. At small angles, however, these two
contributions lead to a new peak, much larger than the original peak
at large angles in the full result. The scalar and vector
contributions show peaks at small angles, and a large peak at large
angles. These results already show that there are large interference
effects present between the various invariant amplitudes. There is
no straightforward and intuitive explanation available for why these
contributions interfere in such a way.

\begin{figure}[ht]
\includegraphics[width=20pc,angle=270]{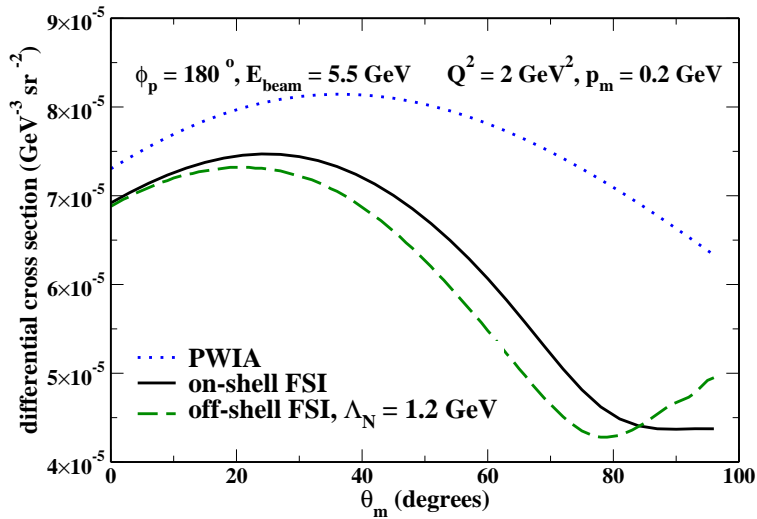}
\caption{The differential cross section for a beam energy of $5.5
$ GeV, $Q^2 = 2~{\rm GeV}^2$, $p_{m} = 0.2$ GeV, and $\phi_p = 180^\circ$ is
shown in PWIA (dotted curve), with on-shell FSI (solid curve), and
with on-shell and off-shell FSIs (dashed curve), as a function of
the angle of the missing momentum. } \label{figcsadlowpm}
\end{figure}

In Fig. \ref{figcsadlowpm}, we show the angular distribution of
the differential cross section for a lower missing momentum, $p_m =
0.2 $GeV. The lower missing momentum implies that the limiting
value of $1.3$ GeV lab energy for the $NN$ system is reached  at
larger angles than for $p_m = 0.4$ GeV. While the PWIA curve here
is very similar in shape to the curve at the higher missing momentum
value, the FSI curve looks rather different now: instead of a fairly
sharp peak around $\theta_m = 70^\circ$, we now observe a broad, shallow
dip at larger angles. At lower missing momenta, the FSIs lead to a
reduction in the cross section. Part of this strength is
redistributed to larger missing momenta, as discussed above. Also
including the off-shell FSIs has no effect at very small angles,
roughly below $20^\circ$, and then tends to shift the overall result
towards lower angles.

\begin{figure}[ht]
\includegraphics[width=20pc,angle=270]{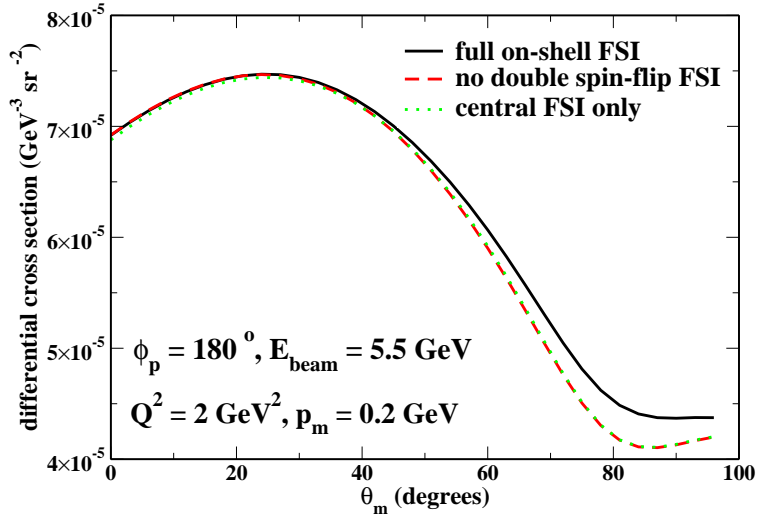}
\caption{The differential cross section for a beam energy of $5.5
$ GeV, $Q^2 = 2~{\rm GeV}^2$, $p_{m} = 0.2$ GeV, and $\phi_p = 180^\circ$ is
shown calculated with on-shell FSI, as a function of the angle of
the missing momentum. The solid line shows the result calculated
with the full $NN$ scattering amplitude, the dashed line shows the
result without the double-spin-flip terms of the $NN$ scattering
amplitudes, and the dotted line shows the result with the central
$NN$ amplitude only. } \label{figcsadlowpmspin}
\end{figure}

In Fig. \ref{figcsadlowpmspin}, the influence of the different
kinds of FSIs is shown at low missing momentum. Switching off the
double-spin-flip contribution leads to a small reduction in the
cross section for medium and large angles, roughly $5\%$ in the
region of the shallow dip. Switching off the single-spin-flip term,
too, changes practically nothing. The FSIs in this kinematic region
are overall smaller than for higher missing momenta. The influence
of spin-dependent FSIs is smaller here, too. However, it is
interesting to note that the double-spin-flip terms are actually
more relevant here than the single-spin-flip terms. We have observed
this already when discussing the momentum distributions shown in
Fig. \ref{figcsmdspin}.

\subsection{Asymmetries}

\begin{figure}[ht]
\includegraphics[height=4in,angle=270]{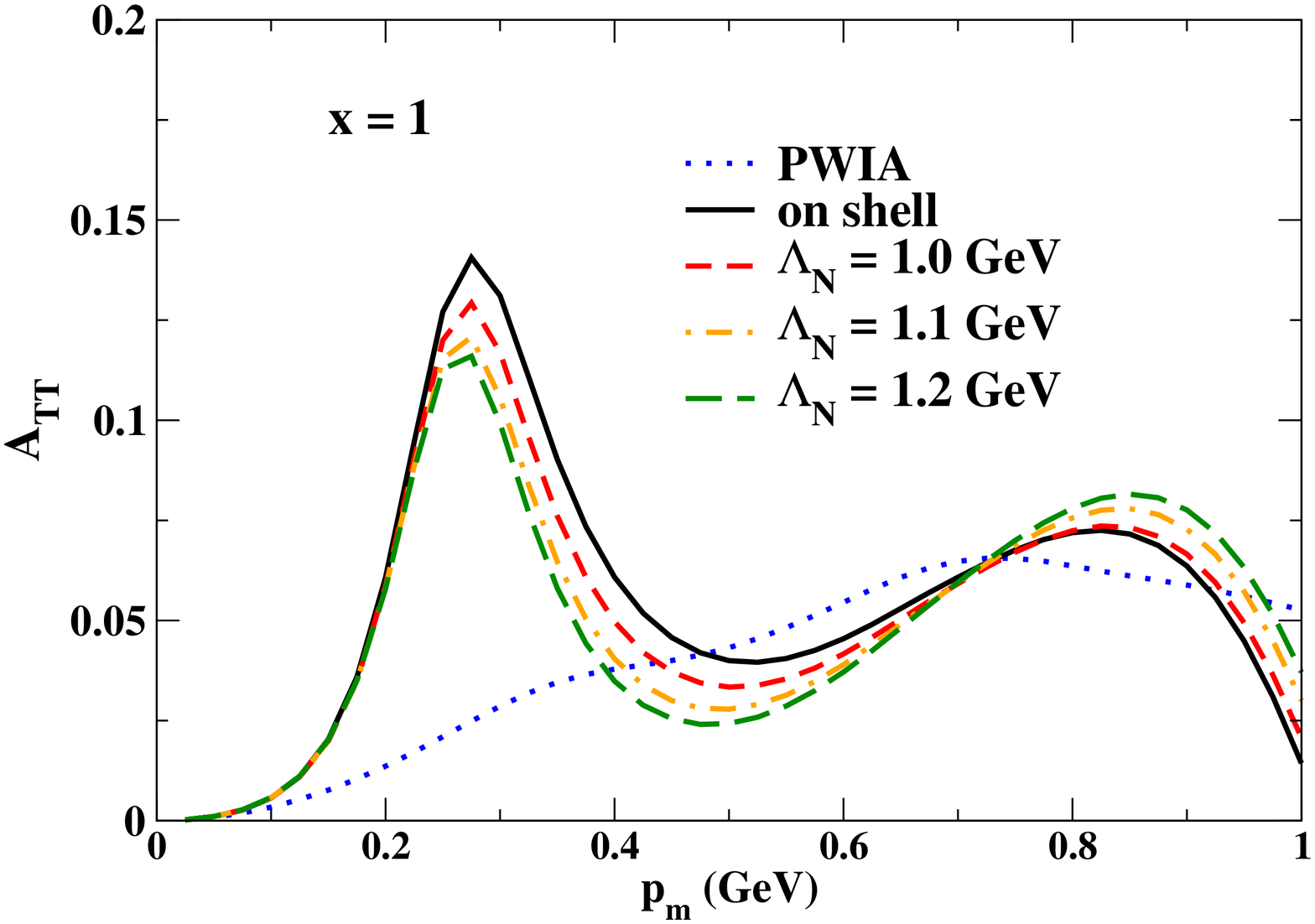}
\includegraphics[height=4in,angle=270]{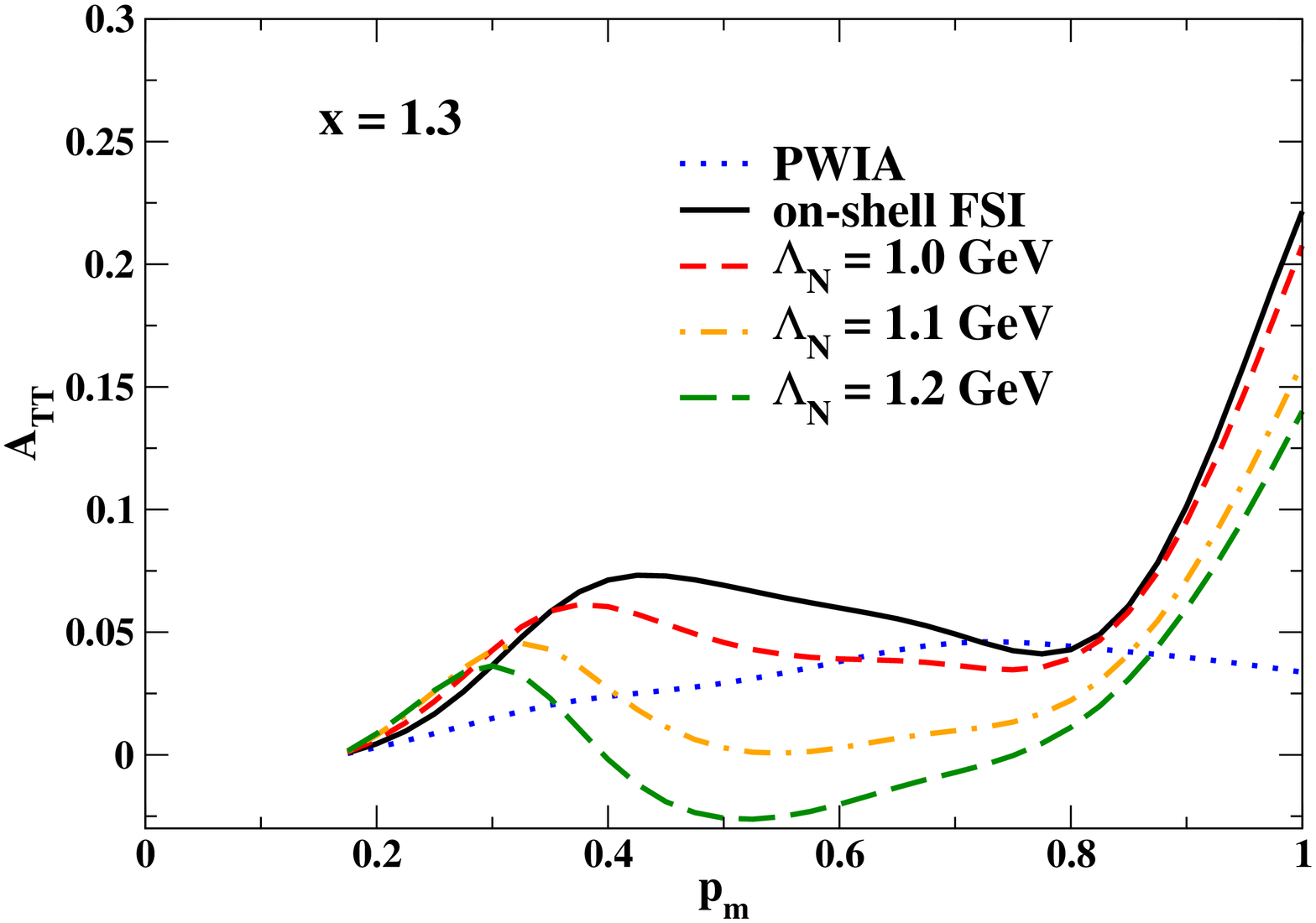}
\caption{Top panel:
the asymmetry $A_{TT}$ is shown for $x_{Bj} = 1$ in PWIA (dotted line),
with on-shell FSI (solid line),
and with off-shell FSI for various values of the cutoff  - $\Lambda_N = 1.0$ GeV (short-dashed line),
$\Lambda_N = 1.1$ GeV (dash - dotted line), and $\Lambda_N = 1.2$ GeV (long-dashed line).
Bottom panel: the asymmetry $A_{TT}$ is shown for $x_{Bj} = 1.3$.
The curves have the same meaning as in the top panel.} \label{figrtt}
\end{figure}

Figure \ref{figrtt} shows the interference asymmetry $A_{TT}$ for $Q^2=2.0$ GeV and $E_{beam}=5.5$ GeV. The upper panel is for $x_{Bj}=1$ and the lower for $x_{Bj}=1.3$.

In both cases, the FSI result in a substantial change from the PWIA result, both with respect to size and shape of the asymmetry.
For $x_{Bj}=1$, the variation with the cutoff mass for the three values shown here is small, but it is somewhat larger for $x_{Bj}=1.3$
reducing the asymmetry almost to zero for $\Lambda_N = 1.2$ GeV around $p_m = 0.5$ GeV.

\begin{figure}[ht]
\includegraphics[height=4in,angle=270]{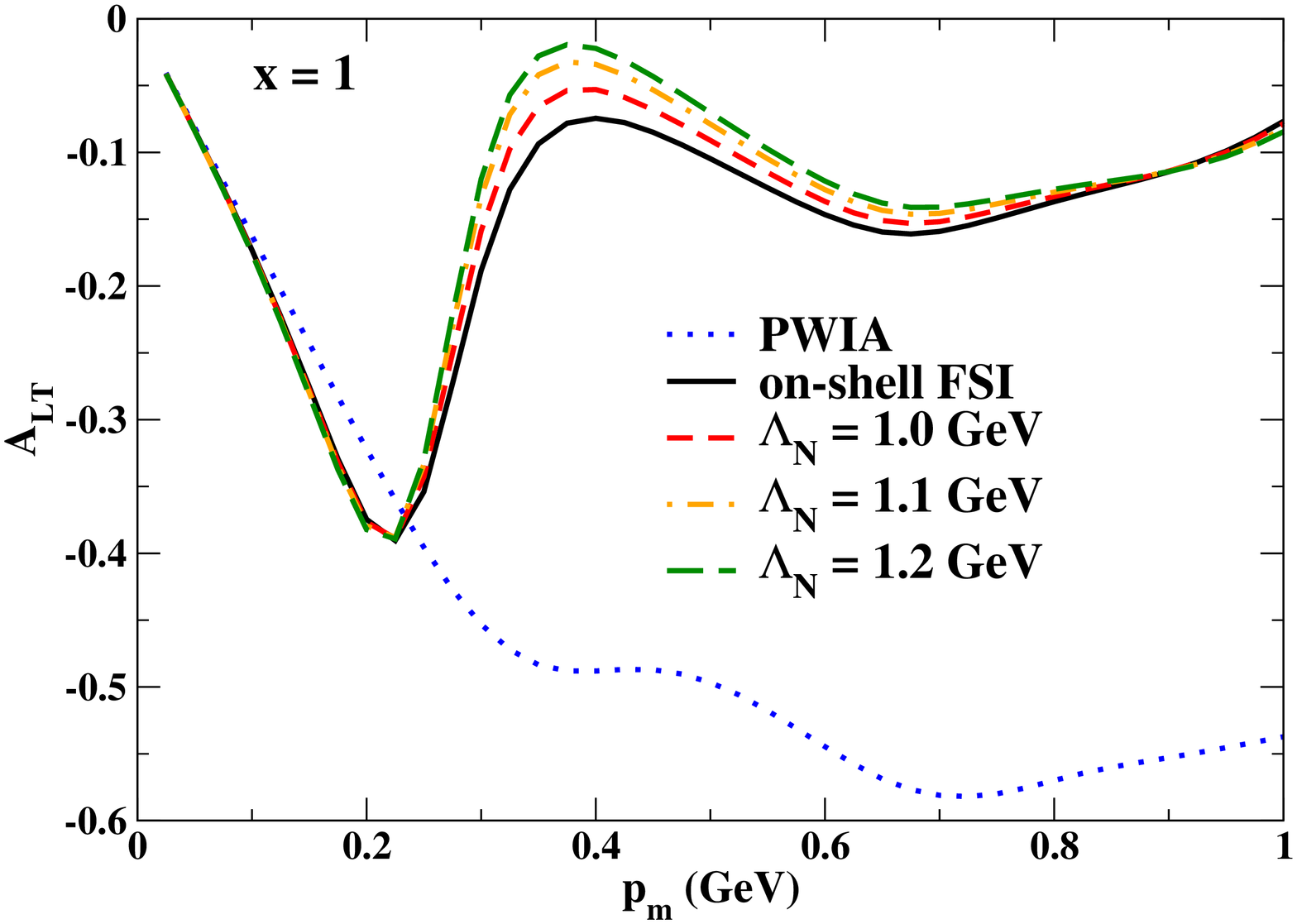}
\includegraphics[height=4in,angle=270]{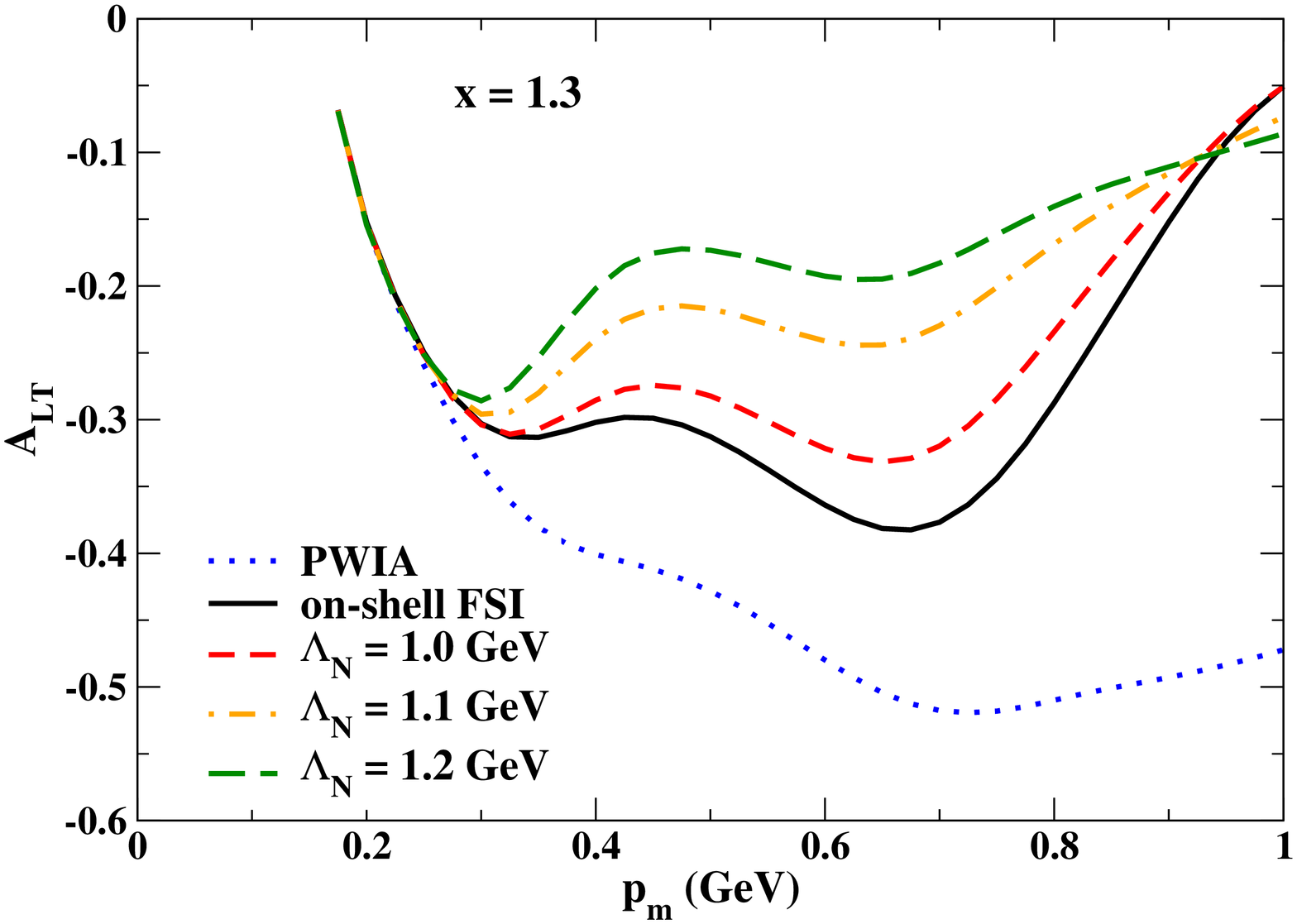}
\caption{Top panel:
the asymmetry $A_{LT}$ is shown for $x_{Bj} = 1$ in PWIA (dotted line),
with on-shell FSI (solid line),
and with off-shell FSI for various values of the cutoff  - $\Lambda_N = 1.0$ GeV (short-dashed line),
$\Lambda_N = 1.1$ GeV (dash - dotted line), and $\Lambda_N = 1.2$ GeV (long-dashed line).
Bottom panel: the asymmetry $A_{LT}$ is shown for $x_{Bj} = 1.3$.
The curves have the same meaning as in the top panel. } \label{figrlt}
\end{figure}
Figure \ref{figrlt} shows results for the interference asymmetry $A_{LT}$ for the same kinematics as in the previous figure. Again the sensitivity to final state interactions is substantial. Sensitivity to off-shell contributions is relatively modest at $x_{Bj}=1$ but is much larger at $x_{Bj}=1.3$.

\begin{figure}[ht]
\includegraphics[height=4in,angle=270]{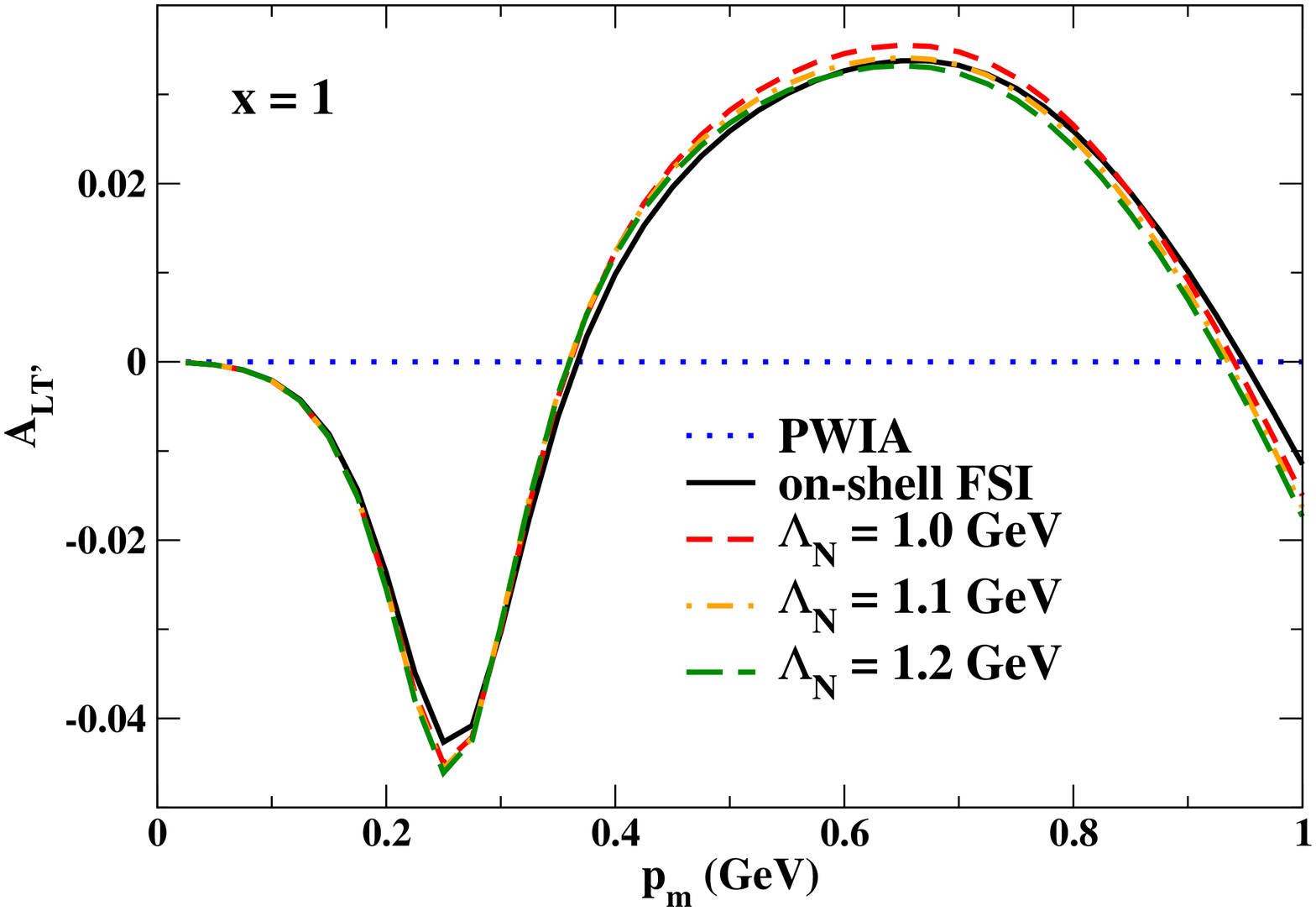}
\includegraphics[height=4in,angle=270]{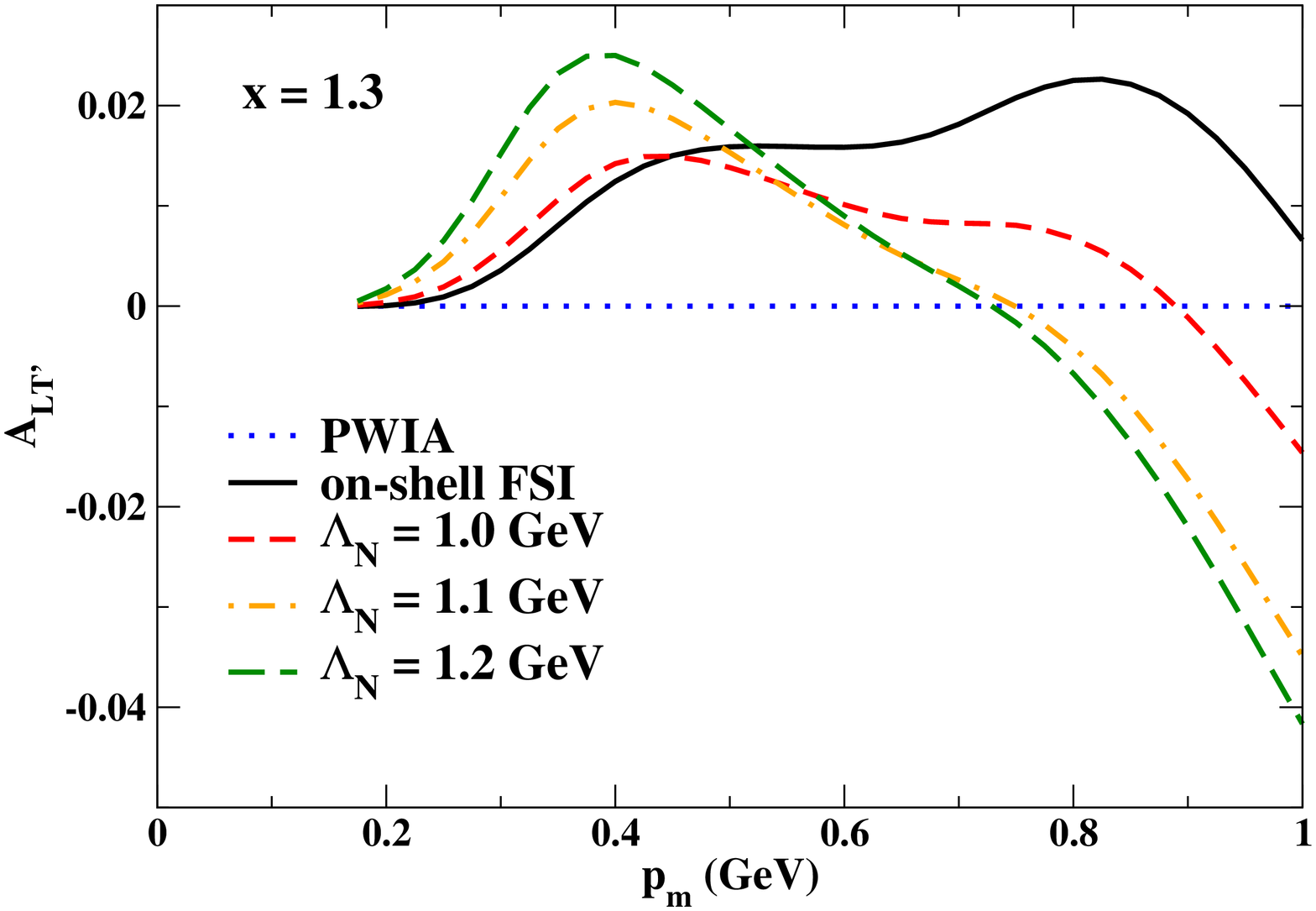}
\caption{Top panel:
the asymmetry $A_{LT'}$ is shown for $x_{Bj} = 1$ in PWIA (dotted line),
with on-shell FSI (solid line),
and with off-shell FSI for various values of the cutoff  - $\Lambda_N = 1.0$ GeV (short-dashed line),
$\Lambda_N = 1.1$ GeV (dash - dotted line), and $\Lambda_N = 1.2$ GeV (long-dashed line).
Bottom panel: the asymmetry $A_{LT'}$ is shown for $x_{Bj} = 1.3$.
The curves have the same meaning as in the top panel. } \label{figrltp}
\end{figure}

The single-spin asymmetry $A_{LT'}$ is shown in Fig. \ref{figrltp} for the same kinematics.  This asymmetry is identically zero in the PWIA, so the FSI are responsible for any asymmetry. A more detailed discussion of FSI effects is given below for kinematics relevant
to a recent experiment. Here, in these kinematics, we focus on comparing the behavior of the three asymmetries, in the same kinematics.

At $x_{Bj}=1$ there is very little sensitivity to off-shell contributions, but it is large at $x_{Bj}=1.3$.
As stated above, this behavior is expected as $x_{Bj} = 1 $ corresponds to quasi-free, i.e. on-shell,
kinematics, while $x_{Bj} = 1.3$ probes nucleons that are much more off-shell. From the plots
for $x_{Bj} = 1.3$, one can see clearly that the off-shell contribution to the NN scattering amplitude
introduces a certain amount of ambiguity, especially at medium to high missing momenta. While great progress
has been made on the experimental side, with measurements at very high missing momenta, there clearly are some
theoretical uncertainties in these kinematic regions.

\begin{figure}[ht]
\includegraphics[width=20pc,angle=270]{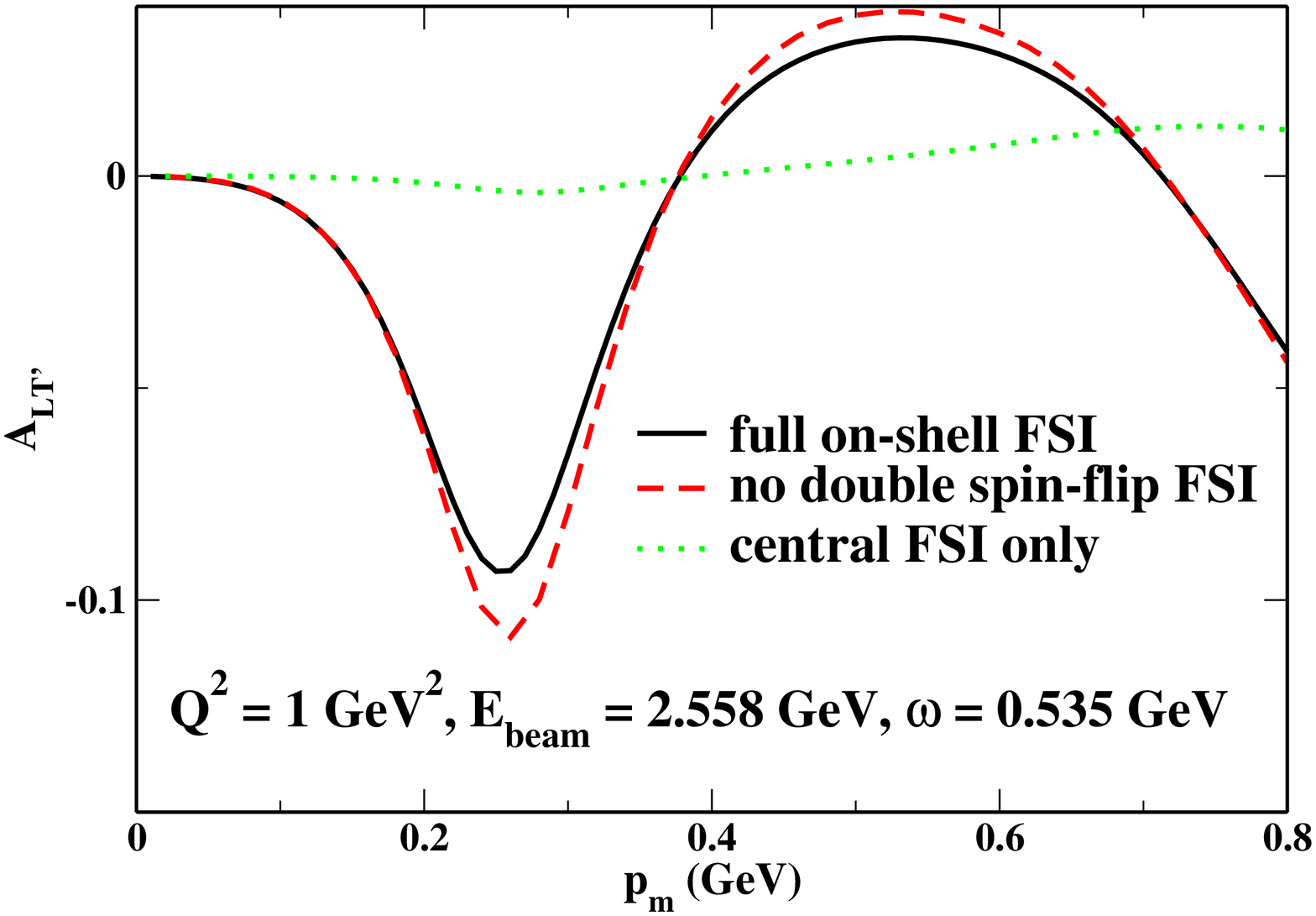}
\caption{The $LT'$ asymmetry for a beam energy of $2.558$ GeV and
four-momentum transfer $Q^2 = 1~{\rm GeV}^2$ is shown calculated with
on-shell FSI, as a function of the  missing momentum. The solid line
shows the result calculated with the full $NN$ scattering amplitude,
the dashed line shows the result without the double-spin-flip terms
of the $NN$ scattering amplitudes, and the dotted line shows the
result with the central $NN$ amplitude only. } \label{figatlpspin}
\end{figure}

We are in the fortunate situation that the asymmetry $A_{LT'}$ has
been measured over a wide range of kinematics, from very low
four-momentum transfers $Q^2 \approx 0.2~{\rm GeV}^2$ up to medium values
of $Q^2 \approx 1 ~{\rm GeV}^2$ \cite{jerrygexp}. The data are currently under analysis.
In this range, the proton-neutron
scattering amplitudes from SAID are available, so there are no
limits to our ability to calculate for these kinematics.

Here, we discuss our results for two representative kinematics: $Q^2
= 0.5 ~{\rm GeV}^2$, and $Q^2 = 1.0 ~{\rm GeV}^2$. In both cases, we assume a beam
energy of $2.558$ GeV. Fig. \ref{figatlpspin} shows our results for
$Q^2 = 1 ~{\rm GeV}^2$. The PWIA result is zero, and not shown on the plot.
The full, on-shell FSI result starts out negatively, dips around
$p_m \approx 0.25$ GeV, and then increases and changes sign around
$p_m \approx 0.38$ GeV. Then, the asymmetry peaks around $p_m
\approx 0.5$ GeV, and then decreases and changes sign again. The
second zero of the asymmetry occurs at $p_m \approx 0.7$ GeV. One
can see that the double-spin-flip contributions to the FSI are not
very relevant: they just lead to small modifications in the dip and
peak regions. The spin-orbit (i.e. the single-spin-flip)
contribution is very important, though. Switching it off so that
only the central FSI remains leads to a completely different
picture: the asymmetry is tiny, and remains positive for large
missing momenta.

\begin{figure}[ht]
\includegraphics[width=20pc,angle=270]{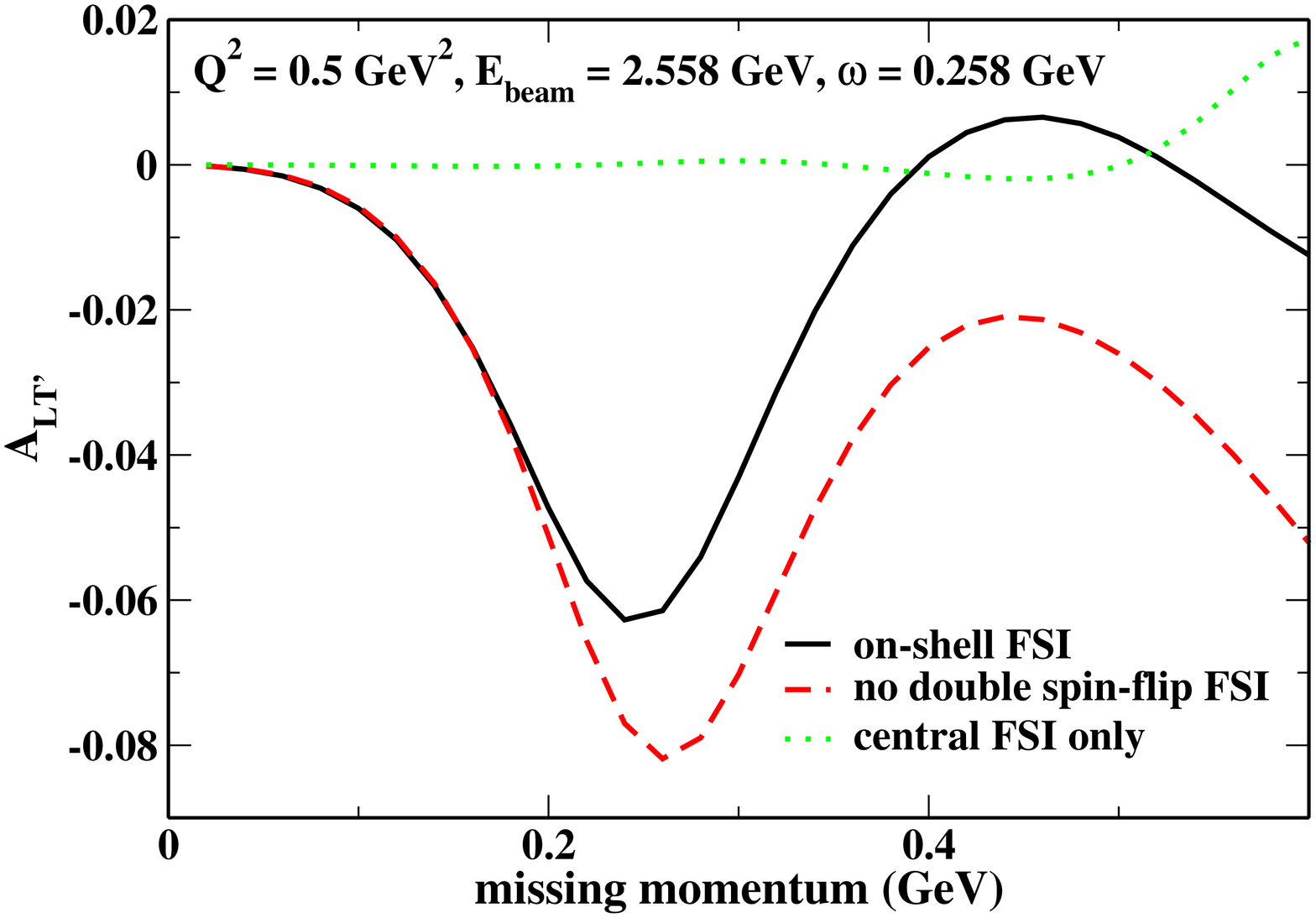}
\caption{The $LT'$ asymmetry for a beam energy of $2.558$ GeV and
four-momentum transfer $Q^2 = 0.5 ~{\rm GeV}^2$ is shown calculated with
on-shell FSI, as a function of the  missing momentum. The solid line
shows the result calculated with the full $NN$ scattering amplitude,
the dashed line shows the result without the double-spin-flip terms
of the $NN$ scattering amplitudes, and the dotted line shows the
result with the central $NN$ amplitude only. }
\label{figatlploqspin}
\end{figure}

In Fig. \ref{figatlploqspin}, we show the corresponding results for the spin-dependence
of the FSIs for lower four-momentum transfer, $Q^2 = 0.5 ~{\rm GeV}^2$. At these kinematics,
we do expect the influence of meson exchange currents and isobar states to be relevant.
These effects are not included in the present calculation. However,
the FSIs are crucial for $A_{LT'}$,
and we can investigate them within our model.
The full calculations are qualitatively very
similar at both $Q^2$ values, showing a negative dip
around $p_m \approx 0.25$ GeV and then an increase
into positive values, with a peak around $p_m \approx 0.5$ GeV.
A very interesting difference, however, is
the size of the contribution of the double-spin-flip terms to the FSI.
While their influence is small, almost negligible,
at the higher $Q^2$ value, it is quite significant for the low $Q^2$ value:
the double-spin-flip terms
serve to partially fill in the negative dip, and are also responsible for pushing the asymmetry
back towards positive values.

\subsection{FSI Details}

One obvious difference between the calculation presented here and the traditional
Glauber and generalized eikonal approximation (GEA) is the evaluation of the argument
of the nucleon-nucleon scattering amplitude for the FSIs. As described in section \ref{sectheo},
we evaluate the five terms of the NN scattering amplitude(\ref{eqdefnn}) at the values of the Mandelstam variables
$s$ and $t$ computed from the particular kinematics. In Glauber and GEA settings,
one typically finds expressions where the NN scattering amplitude is evaluated assuming a purely
transverse momentum transfer when evaluating $t$, even though the longitudinal momentum transfer
is taken into account in the GEA. This is typically denoted with expressions like $f_{NN} (k_{\perp})$.
With the kinematics variables as defined in Fig. \ref{impulse}, the
Mandelstam $t$ is given by
$t = (M_d - E_{k_2} + \omega -E_{p_1})^2 - (\bf{p_1} + \bf{k_2} - \bf{q})^2$,
whereas assuming a purely transverse momentum transfer implies:
$t_{\perp} = -(\bf{p_{1,\perp}} + \bf{k_{2,\perp}})^2$.
Using this we can define the cm scattering angle as
\begin{equation}
\cos\theta=1+\frac{2t_\perp}{s-4m^2}\label{thetacm_perp}
\end{equation}

\begin{figure}[ht]
\includegraphics[width=20pc,angle=270]{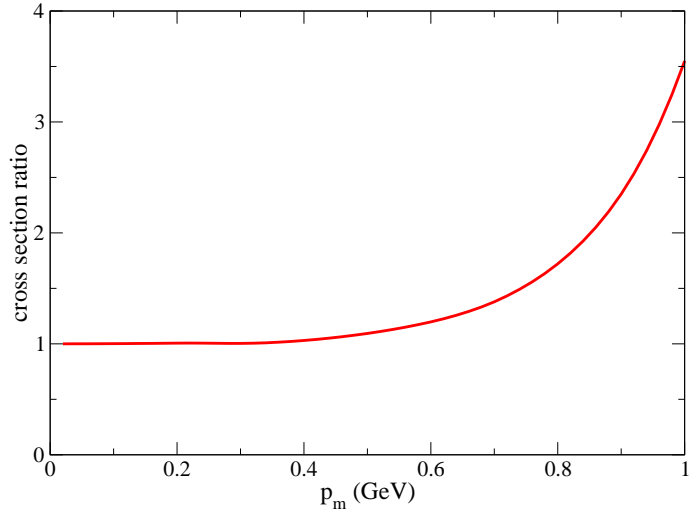}
\caption{The ratio of the differential cross section evaluated with the full $NN$ argument, $t$,
and with the purely transverse momentum, $t_{\perp}$. The kinematics are identical to
Fig. \ref{figcsmomdis}:  beam energy of $5.5
$ GeV, $Q^2 = 2 ~{\rm GeV}^2$, $x_{Bj} = 1$, and $\phi_p = 180^\circ$. The calculation was performed
using on-shell FSI.}
\label{fignnratiocsmd}
\end{figure}

In Fig. \ref{fignnratiocsmd}, we show the ratio of the transverse-momentum approximation to the
full, on-shell calculation.  The kinematics are identical to the kinematics used for
Fig. \ref{figcsmomdis}:  a beam energy of $5.5$ GeV, $Q^2 = 2 ~{\rm GeV}^2$, $x_{Bj} = 1$, and $\phi_p = 180^\circ$.
Up to missing momenta of $0.4$ GeV, the two approximation works well, leading
to small deviations of less than $5\%$. Beyond $0.4$ GeV, the deviation from the full result grows, and for
missing momenta larger than $0.6$ GeV, the quality of the approximation deteriorates quickly.

We have performed the same calculation for the angular distribution shown in Fig. \ref{figcsangdis}.
Here, we found that the approximation does well - the deviations are less than $5\%$ for any angle.
This corresponds to our findings for the momentum distribution.

\begin{figure}[ht]
\includegraphics[width=20pc,angle=270]{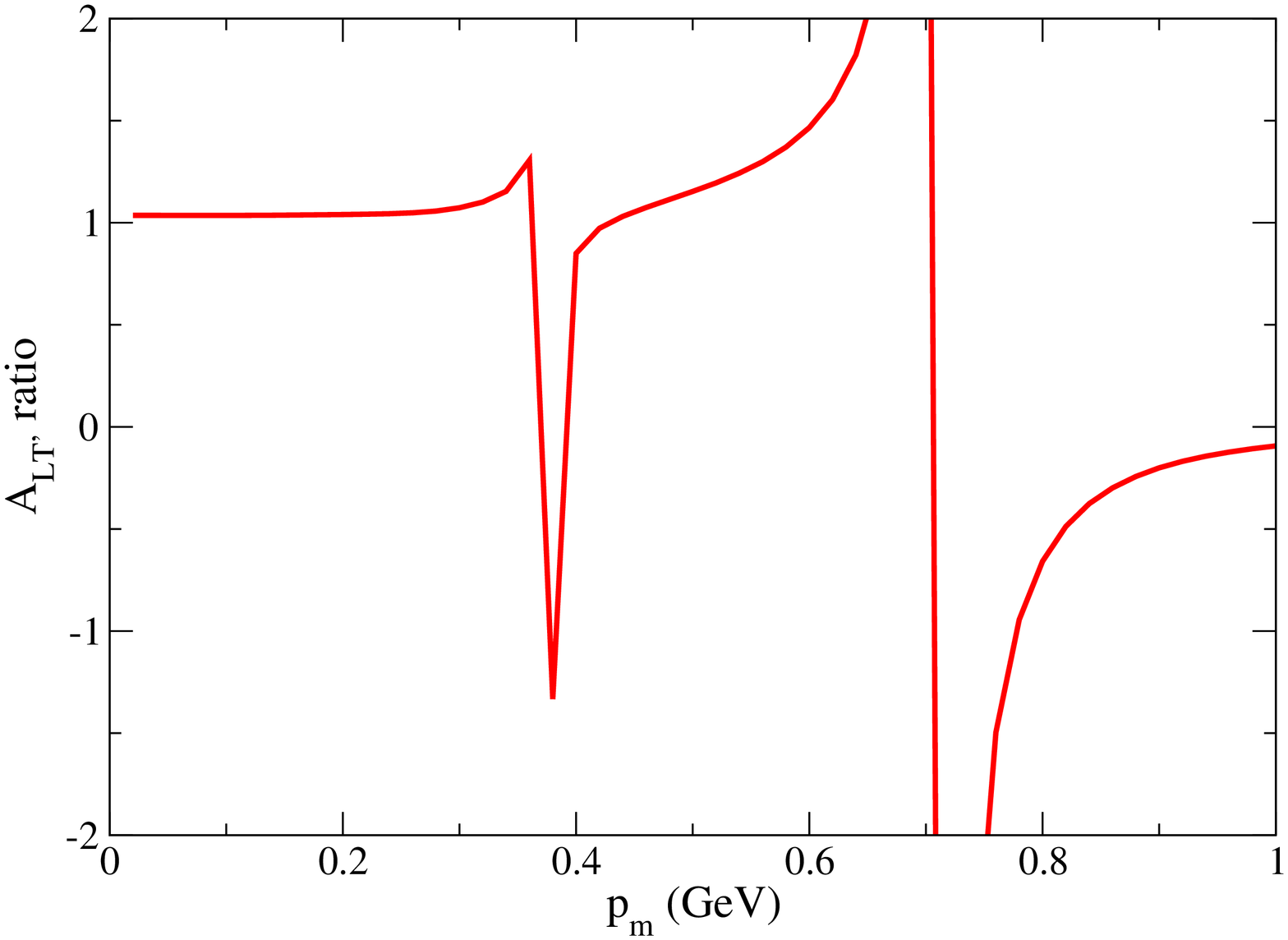}
\caption{The ratio of the asymmetry $A_{LT'}$ evaluated with the full $NN$ argument, $t$,
and with the purely transverse momentum, $t_{\perp}$. The kinematics are identical to
Fig. \ref{figatlpspin}:   The calculations were performed
using on-shell FSI.}
\label{fignnratioaltp}
\end{figure}

As an illustrative example for the other observables discussed in this article, we also show our
results for the asymmetry $A_{LT'}$ in Fig. \ref{fignnratioaltp}. The kinematics correspond to
Fig. \ref{figatlpspin}, we have used a beam energy of $2.558$ GeV and
four-momentum transfer $Q^2 = 1~{\rm GeV}^2$. The spike seen in the ratio around $p_m = 0.4$ GeV stems from the
sign change in the asymmetry, they are not relevant to our discussion. Again, we see that the approximation
is doing well up to missing momenta of roughly $0.4$ GeV. Then, the approximation considerably
overestimate the full result, and for very large missing momenta, $p_m \ge 0.7$ GeV, it even fails to
reproduce the correct sign of the asymmetry. It is interesting to see that the effects of the
approximation are visible even in a quantity that is a ratio of quantities that are both affected by
the approximation.

In conclusion, approximating the argument of the NN scattering amplitude with the popular ``transverse
momentum transfer only'' works well up to missing momenta of $0.4$ GeV for various observables.
For missing momenta higher than that, the approximation becomes questionable.

\section{Summary and Outlook}

In this paper, we have presented a fully relativistic $D(e,e'p)$ calculation in impulse approximation.
We have steered clear
from a number of common, simplifying assumptions.
The only approximation made in this paper is neglecting the negative-energy contributions to the propagator of the off-shell nucleon. These contributions can realistically expected to be
very small compared to the positive-energy contributions.

We have used a parametrization of experimental $NN$ data from SAID to describe the full $pn$ scattering amplitude
for the final state interaction. This leads to certain limits in the kinematics we can access, as these parametrizations
are available only for lab kinetic energies of $1.3$ GeV or less. In our calculations, we have investigated the effects
of the different contributions to the NN scattering amplitude: the central, spin-orbit, and double-spin-flip parts,
using the Saclay formalism to describe the different contributions. Many other fine calculations using the generalized
eikonal approximation \cite{misak,ciofi,genteikonal} or a diagrammatic approach \cite{laget} use the central part only.
While the central part of the amplitude is clearly dominant in almost all observables, the spin-orbit and double-spin-flip
parts do contribute visibly to the cross section in the peak area of the angular distribution,
increasing the peak height roughly by a quarter.
For the out-of-plane asymmetry $A_{LT'}$, which is non-zero only in the presence of FSIs, the spin-orbit part is clearly
the most relevant. Depending on the kinematics, the double-spin-flip can also play a relevant role for this asymmetry.

We also showed the different contributions of the NN amplitude in terms of the five invariant amplitudes. Interestingly,
they are all relevant, and a lot of interference effects contribute to the full result. It is not possible to identify a
single, dominant contribution in this framework for the description of the NN amplitudes.

In the spirit of avoiding all unnecessary approximations, we have used the full argument for the calculation of the
NN scattering amplitudes. In Glauber theory and its variants, one often encounters the assumption of a transverse
momentum transfer only, and this changes the value of Mandelstam $t$. We have investigated the validity of this assumption,
and found that it is a very good approximation for missing momenta up to $0.4$ GeV. For higher missing momenta, the quality
of this approximation deteriorates quickly, and it should probably not be used.

We have also compared the influence of the off-shell FSI contributions to the on-shell FSI contributions. The former
are expected to not be too large, and they require some interpolation of on-shell amplitudes and the introduction of a
regulator function to suppress very far off-shell contributions. The off-shell FSI contributions tend to be negligible to
small for lower missing momenta, $p_m \leq 0.4$ GeV for any observable. Beyond that, their importance varies
depending on the specific kinematics: the off-shell FSI is very small for the momentum distribution calculated for the
quasi-free value $x_{Bj} = 1$, but it is large for $x_{Bj} = 1.3$. This pattern was observed both for the cross section
and the asymmetries $A_{TT}$, $A_{LT}$, and $A_{LT'}$. The size of the off-shell contribution does depend on the chosen
cut-off, with a larger cut-off admitting a sometimes much-larger contribution. The main purpose of showing figures with the
ratios of off-shell calculations with different cut-offs to the on-shell result is to identify ``safe'' kinematics and
observables, where the off-shell FSI contributions are definitely small. In regions where they are relevant, a certain
amount of theoretical uncertainty cannot be avoided, until reliable and believable models of the off-shell NN interaction
at the relevant energies are developed. This is particularly relevant for the interpretation of new data taken at high
missing momentum at Jefferson Lab.

The current calculations will be applied or already have been applied to the forthcoming experimental data from Jefferson Lab \cite{jerrygexp,halladata}, and calculations for the BLAST data from MIT Bates \cite{blast} are planned.
Our calculation also does a nice job of improving the agreement with the low missing momentum data of Ulmer et al. \cite{paulprl},
even though the ``low missing momentum puzzle'' \cite{boeglintrento2005} is not completely resolved.

Logical next steps for enhancing our calculations are the inclusion of meson exchange currents, and isobar states.


{\bf Acknowledgments}:  We thank Paul Ulmer for quickly providing us
with his data from \cite{paulprl} in tabulated form. SJ thanks Charlotte Elster for
interesting discussions. This work was
supported in part by funds provided by the U.S. Department of Energy
(DOE) under cooperative research agreement under No.
DE-AC05-84ER40150 and by the National Science Foundation under
grants No. PHY-0354916 and PHY-0653312.

\appendix
\section{Representations of the $NN$ amplitudes}
\label{appnn}

The invariant functions $\mathcal{F}_i(s,t)$ can be obtained from
scattering data. For example, the helicity amplitudes are defined as
\begin{equation}
\mathcal{M}_{\lambda_1'\lambda_2';\lambda_1\lambda_2}=(\bar{u}_{\lambda_{1}'}
(\bm{p}_1'))_{a}(\bar{u}_{\lambda_2'}(\bm{p}_2'))_{b}M_{ab,cd}(u_{\lambda_1}
(\bm{p}_1))_c(u_{\lambda_2}(\bm{p}_2))_d\label{helicity}
\end{equation}
where $u_{\lambda}(\bm{p})$ is the helicity spinor for helicity
$\lambda$. The helicity matrix elements for $pn$ scattering in the
center-of-momentum frame can be obtained from the program SAID for
laboratory kinetic energies of up to 1.3 GeV.  If the amplitudes are
extracted in units of $fm$, the helicity amplitudes consistent with
the conventions used here are related to the SAID amplitudes by
\begin{equation}
\mathcal{M}_{\lambda_1'\lambda_2';\lambda_1\lambda_2}=-\frac{4\pi\sqrt{s}}{\hbar
c m^2}
\mathcal{M}_{\lambda_1'\lambda_2';\lambda_1\lambda_2}^{SAID}\,.
\end{equation}
Parity, time-reversal and particle interchange symmetries can be
used to show that there are only five independent helicity
amplitudes defined as
\begin{eqnarray}
a&=&\mathcal{M}_{1,1;1,1}\\
b&=&\mathcal{M}_{1,1;1,-1}\\
c&=&\mathcal{M}_{1,-1;1,-1}\\
d&=&\mathcal{M}_{1,1;-1,-1}\\
e&=&\mathcal{M}_{1,-1;-1,1}\,.
\end{eqnarray}
Using (\ref{Fermi}) in (\ref{helicity}) and solving for the
invariant functions gives
\begin{equation}
\begin{pmatrix}
\mathcal{F}_S\\
\mathcal{F}_{V}\ \\
\mathcal{F}_{T}\ \\
\mathcal{F}_{P}\ \\
\mathcal{F}_{A}\ \\
\end{pmatrix}
=\frac{1}{s-4m^{2}}
\begin{pmatrix}
a_{11} & a_{12} & a_{13} & a_{14} & a_{15} \\
a_{21} & a_{22} & a_{23} & a_{24} & a_{25} \\
a_{31} & a_{32} & a_{33} & a_{34} & a_{35} \\
a_{41} & a_{42} & a_{43} & a_{44} & a_{45} \\
a_{51} & a_{52} & a_{53} & a_{54} & a_{55} \\
\end{pmatrix}
\begin{pmatrix}
a \\
b \\
c \\
d \\
e \\
\end{pmatrix}
\label{HetoInv}
\end{equation}
where
\begin{eqnarray}
a_{11}&=&-a_{24}=a_{25}=-2a_{31}=a_{41}=-a_{54}=-a_{55}=-\frac{2m^{4}}{s}\\
a_{14}&=&-a_{15}=-a_{21}=-2a_{34}=2a_{35}=a_{44}=-a_{51}= \frac{2m^4}{s}-m^2\\  a_{12}&=&\frac{m[8m^2-(3+\cos\theta)s]}{\sqrt{s}\sin\theta}\\
a_{13}&=&\frac{m^2[2m^2(1+\cos\theta)-s(3+\cos\theta)]}{s(1+\cos\theta)}
\\
a_{22}&=&\frac{4m^{2}(1+\cos\theta)}{\sqrt{s}\sin\theta}\\
a_{23}&=&\frac{2m^{2}[m^{2}(1+\cos\theta)+s]}{s(1+\cos\theta)}\\
a_{32}&=&-\frac{m\sqrt{s}(1-\cos\theta)}{2\sin\theta}\\
a_{33}&=&-\frac{m^2[m^2(1+\cos\theta)+s(1-\cos\theta)]}{s(1+\cos\theta)}\\
a_{42}&=&-\frac{m[8m^{2}+s(3+\cos\theta)]}{\sqrt{s}\sin\theta}\\
a_{43}&=&\frac{m^{2}[2m^2(1+\cos\theta)-s(3+\cos\theta)]}{s(1+\cos\theta)}\\
a_{45}&=&-\frac{m^{2}[2m^2(1-\cos\theta)+s(7+\cos\theta)]}{s(1-\cos\theta)}\\ a_{52}&=&-\frac{4m^{2}(1-\cos\theta)}{\sqrt{s}\sin\theta}\\
a_{53}&=&\frac{2m^{2}[m^{2}(1+\cos\theta)-s]}{s(1+\cos\theta)}\,.
\end{eqnarray}

The Saclay amplitudes are given in terms of the helicity amplitudes
by
\begin{equation}
\begin{pmatrix}
a_s \\
b_s \\
c_s \\
d_s \\
e_s \\
\end{pmatrix}
= \frac{1}{2\sqrt{2}}
\begin{pmatrix}
\cos\theta & -4\sin\theta &\cos\theta &\cos\theta & -\cos\theta \\
1 & 0 & 1 & -1 & 1 \\
-1 & 0 & 1 & 1 & 1 \\
1 & 0 & -1 & 1 & 1 \\
-i \sin\theta & -4i\cos\theta &-i \sin\theta &-i \sin\theta &i \sin\theta \\
\end{pmatrix}
\begin{pmatrix}
a \\
b \\
c \\
d \\
e \\
\end{pmatrix}\,.
\label{HetoSa}
\end{equation}
Equations (\ref{HetoInv}) and (\ref{HetoSa}) can then be used to
obtain the transformation from the Saclay amplitudes to the Fermi
invariant functions.

\end{document}